\documentclass[a4paper,11pt]{article}

\pdfoutput=1

\usepackage{jheppub}
\usepackage{amsmath,amssymb,amsfonts,mathtools,amsthm}
\usepackage{xcolor}
\usepackage{graphicx}
\usepackage{caption}
\usepackage{subcaption}
\usepackage{floatrow}
\usepackage{physics}
\usepackage{tensor}
\usepackage{bbm}
\usepackage{url}
\usepackage{booktabs}
\usepackage{stackrel}
\usepackage{hyperref}
\usepackage[toc,page]{appendix}
\usepackage{comment}
\usepackage{lipsum}
\usepackage{amssymb}
 \usepackage{mathrsfs}

\usepackage{tikz}
\usetikzlibrary{quantikz, matrix, decorations.markings, calc, shapes, 
  decorations.pathmorphing, patterns, decorations.pathreplacing, positioning}

\makeatletter
\def\@fpheader{\relax}
\makeatother

\theoremstyle{plain}

\preprint{UT-WI-22-2026}
\title{Aspects of Carrollian field theory from holography}

\author{Hare Krishna and Vaishnavi Patil}

\affiliation{ Weinberg Institute, Department of Physics, University of Texas at Austin, Austin, TX 78712, USA}

\emailAdd{hkrishna.phy@gmail.com}\emailAdd{vaishnavi.patil@utexas.edu}

\begin{document}
% \abstract{In this article, we study the holographic principle for 3-dimensional asymptotically flat spacetimes. The gravitational physics in 3d bulk is conjectured to be dual to the 2-dimensional Carrollian/BMS field theory. These field theories have previously been understood as ultra-relativistic contractions of relativistic 2D conformal field theories. Using this contraction, the Virasoro algebra becomes the 2d conformal Carrollian (equivalently, BMS) algebra and the Brown-Henneaux central charges contract to $c_L=0,\;c_M=\frac{3}{G}$. We derive these central charges using holography. We first obtain the holographic quasilocal stress tensor for asymptotically flat spacetimes with Bondi mass and angular momentum. The holographic stress tensor transforms under BMS transformations with a non-homogeneous term, i.e. ``BMS Schwarzian", from which we extract the central charges and find exact agreement with the algebraic approach. Furthermore, we obtain the action for the boundary BMS Schwarzian modes. The stress tensor correlators satisfy the Ward identity, and using this, we established the IR triangle of asymptotic symmetries, soft theorems, and memory effects.}
\abstract{We study holography for three-dimensional asymptotically flat spacetimes in which the bulk gravitational dynamics is conjectured to be dual to a two-dimensional Carrollian (equivalently BMS) conformal field theory. Such theories are known to arise as ultrarelativistic ($c\rightarrow0$) contractions of relativistic 2d conformal field theories. Under this contraction, the Virasoro algebra becomes the conformal Carrollian algebra and the Brown-Henneaux central charges become $c_L=0$, $c_M=\frac{3}{G}$. In this article, we recover these central charges directly from holography. We first construct the holographic quasilocal stress tensor for asymptotically flat spacetimes carrying Bondi mass and angular momentum. Under BMS$_3$ transformations, the stress tensor acquires an inhomogeneous term, the ``BMS Schwarzian", and extracting the central charges from it reproduces the algebraic result exactly. We then obtain the action for the boundary BMS Schwarzian modes. Finally, we show that the holographic stress-tensor correlators satisfy the expected Ward identity, while stress-tensor conservation obeys flux-balance laws.}

\maketitle
\section{Introduction and Summary of Results}
The AdS/CFT correspondence provides a precise dictionary between gravitational dynamics in asymptotically anti-de Sitter (AdS) spacetimes and the dynamics of a conformal field theory living on the boundary \cite{Maldacena:1997re,Witten:1998qj,Gubser:1998bc}. It furnishes our most tractable non-perturbative definition of quantum gravity in such backgrounds. Two structural features of gravity underlie the naturalness of this construction: the Hamiltonian is a pure boundary term, and gauge-invariant
observables admit no local definition in the bulk but must instead be anchored to the boundary. Neither feature relies on a negative cosmological constant. This motivates the search for holographic descriptions of asymptotically flat and de Sitter spacetimes as well, even in the absence of a string-theoretic embedding to guide the construction.\\

In this article, we pursue a holographic description for three-dimensional asymptotically flat spacetimes (AFS). Three dimensions provide a controlled setting: the bulk has no propagating gravitons \cite{Deser:1983tn,Witten:1988hc}, all the dynamics is boundary dynamics, but still the metric for AFS is non-trivial with Bondi mass and angular momentum. AFS in 3d has the asymptotic symmetry algebra BMS$_3$ \cite{Barnich:2006av,Barnich:2012aw}, an infinite-dimensional extension of the Poincar\'e algebra similar to the BMS$_4$ in 4d \cite{sachs_asymptotic_1962,Bondi1962,Barnich:2009se,Barnich:2010eb}.  Many aspects of this problem have been studied from the complementary perspectives of asymptotic
symmetries, covariant phase space formalism, soft theorems, and memory effects   \cite{Strominger:2014pwa,Strominger:2017zoo,deBoer:2003vf,He:2014laa,Cachazo:2014fwa}. Our approach here is a little different. \\

We adapt the holographic renormalization of the gravitational action developed for AdS by Balasubramanian and Kraus \cite{Balasubramanian:1999re}: the on-shell action, supplemented by boundary counterterms constructed from intrinsic boundary data, is varied with respect to that boundary metric to define a renormalized quasi-local stress tensor, whose
transformation properties then encode both the asymptotic symmetry algebra and its central extension. In AdS$_3$ this procedure yields a boundary stress tensor transforming in two copies of the Virasoro algebra with the Brown-Henneaux central charge $c = 3\ell/2G$ \cite{Brown:1986nw}. The flat-space case is not obtained by naively sending $\ell \to \infty$ in that
result. Both the counterterms and the boundary geometry must be reconsidered, since null infinity carries a degenerate (Carrollian) rather than Lorentzian structure. \\

% Furthermore, the asymptotic boundary of AFS consists of past/future null infinity $\mathscr{I}^{\pm}$, along with space-like $i^0$ and time-like infinities $i^{\pm}$. The boundary is a union of all of these. But there are set of conservation equation which needs to be matched as well \cite{Strominger:2014pwa,Strominger:2017zoo}. Therefore, we need to start from foliation whose boundary limit will match with the structure that we are interested in. For an example, if we wanna construct a quasilocal stress tensor on null inifinity we may foliate the spacetime with null like hypersurface \cite{Ciambelli:2025mex} or time like hypersurface having induced carroll structure \cite{,Ashtekar:1981bq,Freidel:2022vjq,Freidel:2024emv,Bhambure:2024ftz} which gets idenitified with Carroll structure at null infinity when pushed to the boundary. Similarly, if wanna construct the holographic dictionary at spacelike infinity, then we may need to foliate the spacetime with dS$_3$ slicing \cite{deBoer:2003vf,BeigSchmidt1982,AshtekarHansen1978} and its asymptotic limit gives a construction near $i^0$. One can further do the foliation near time like infinity by foliating with hyperbolic slicing. A complete and unified perspective has been presented in \cite{Compere:2023qoa}. All these 5 set of inifinities $(\mathscr{I}^{\pm},i^{\pm},i^0)$ share a single BMS algebra of asymptotic symmetries and associated charges. In this article, we will construct stress tensor and holographic dictionary mainly on the null infinity and leave the analogus construction on space and time infinity for future work.

The conformal boundary of an asymptotically flat spacetime is not a single connected component. It decomposes into future and past null infinity $\mathscr{I}^{\pm}$, spatial infinity $i^{0}$, and future and past timelike infinity $i^{\pm}$. But it is not just a disjoint union of these regions since the conservation laws associated with asymptotic symmetries relate incoming to outgoing data. The fields are glued across $i^{0}$ by antipodal matching conditions \cite{Strominger:2014pwa,Strominger:2017zoo}.
This has a practical consequence for holography. Because each region carries a different asymptotic geometry, it is difficult to construct boundary observables in a region-independent way. Alternatively, the bulk foliation can be chosen so that its boundary limit reproduces the structure one wishes to probe. To construct a quasi-local stress tensor or holographic dictionary at null infinity, one may foliate by null hypersurfaces \cite{Ciambelli:2025mex}, or by a family of timelike hypersurfaces whose induced geometry degenerates to a Carrollian structure as they are pushed to $\mathscr{I}^\pm$ \cite{Freidel:2022vjq,Freidel:2024emv,Bhambure:2024ftz}. For a dictionary at spatial infinity, one instead foliates by de-Sitter slicing, which in four bulk dimensions is made of dS$_{3}$ slices \cite{deBoer:2003vf,BeigSchmidt1982,AshtekarHansen1978}.
The construction near $i^{\pm}$ proceeds along the same lines, with hyperboloids giving the hyperbolic $H_{3}$ slices. A unified treatment of these regions has been given in \cite{Compere:2023qoa}. The asymptotic symmetries of these regions are not independent. At each of $\mathscr{I}^{\pm}$ one finds a copy of BMS$_{4}$, and the matching conditions across $i^{0}$ reduce these to a single diagonal BMS$_{4}$ acting on the full scattering problem, with the corresponding charges conserved. Furthermore, with massive and massless particle scattering, it has been shown by Comp\'{e}re et al. \cite{Compere:2023qoa} that these 5 infinities share a single BMS symmetry group and sets of conserved charges. We believe a similar construction exists for AFS$_3$ as well. In this article, we construct the boundary stress tensor and holographic dictionary at null infinity, leaving the analogous constructions at spatial and timelike infinity for future work.\\

From the dual field theory side, the QFT is considerably less developed. Defining a quantum field theory on $\mathscr{I}^{\pm}$ is more complicated by the presence of a degenerate metric and a nowhere-vanishing vector field in its kernel. The theories that do appear are called Carrollian/BMS field theories (or celestial CFTs if defined only on the spatial sphere $S^{d-2}$). Two-dimensional BMS field theory (BMSFT$_{2}$) is supposed to encode the gravitational dynamics in AFS$_{3}$. Two strategies have been pursued so far. The first obtains these theories as ultra-relativistic contractions of relativistic QFTs, sending the effective speed of light
$c \rightarrow 0$ so that the Poincar\'e algebra contracts to the Carroll algebra \cite{Duval:2014uoa, Ruzziconi:2026bix,Bagchi:2025vri,Alday:2024yyj,Kulkarni:2025qcx,Lipstein:2025jfj}. This is also equivalent to taking $\ell_{\rm AdS} \rightarrow \infty$ limit of AdS correlators which has been studied in \cite{Alday:2024yyj,Lipstein:2025jfj,Kulkarni:2025qcx}. In such a limit, the Brown-Henneaux central charge $c=\bar{c} = 3\ell/2G_{N}$
\cite{Brown:1986nw} diverges, and we need to scale the Virasoro generators appropriately to get the finite central charges $c_L=0, c_M=\frac{3}{G}$ \cite{Barnich:2006av}.  Such limits do capture genuine features of flat-space holography, but as
emphasised by Susskind \cite{Susskind:1998vk} they are delicate
from the boundary perspective. For traditional AdS$_5$ and super Yang-Mills duality, this is equivalent to $N \rightarrow \infty$ but keeping $g^2_{\text{YM}}$ fixed.  The parametric limit alone does not
suffice. A flat-space scattering process must be localized in a region $\ell_{s} \ll R \ll \ell$, and is therefore built from states of energy $E \gg 1/\ell$. Then the relevant boundary operators carry dimensions $\Delta \sim E\ell \rightarrow \infty$ and the S-matrix is extracted not from correlators of fixed operators but from a particular asymptotic corner of a family of correlators whose external data moves with the limit (see \cite{Giddings:1999qu,Polchinski:1999ry} for the AdS case). The second strategy is to use asymptotic symmetry algebra, its
representations and correlators intrinsically
\cite{Barnich:2009se,Bagchi:2009pe,Bagchi:2010zz,Hao:2021urq,Bagchi:2022eav}. To avoid these subtleties, we pursue the holographic method directly in AFS without taking any limit.\\

We study three-dimensional asymptotically flat spacetimes in Bondi gauge, characterised by a mass aspect $M$ and an angular momentum aspect $N$. The constraint components of the vacuum Einstein equations fix the retarded-time dependence of these aspects entirely, leaving two arbitrary functions $M(\phi)$ and $N(\phi)$. This free data parametrises the space of solutions. Adapting the formalism of quasi-local stress tensors at null boundaries
\cite{Chandrasekaran:2021hxc,Bhambure:2024ftz,Freidel:2022vjq,Freidel:2024emv,Ciambelli:2025mex}, we construct the boundary stress tensor for AFS at $\mathscr{I}^{+}$. Its components are built from $M(\phi)$ and $N(\phi)$. Although three-dimensional gravity propagates no bulk gravitons, it does have the boundary degrees of freedom called BMS Schwarzian \cite{Merbis:2019wgk,Cotler:2024cia}. This is the flat-space
counterpart of the Schwarzian theory in AdS$_3$. Our results are as follows.

 \begin{itemize}
    \item For an asymptotically flat spacetime (AFS) having the metric $ds^2 = M(\phi)du^2  -2 dudr +2 \Big[N(\phi) + \frac{u}{2} \partial_\phi M(\phi) \Big]dud\phi + r^2 d\phi^2 $, the components of holographic stress tensor at $\mathscr{I}^+$ can be computed as
    \begin{equation}
    \begin{split}
        & T_u{}^u = -\frac{1}{8\pi G_N} \frac{M(\phi)}{2\, r} + O\left(\frac{1}{r^2}\right) \\
        & T_\phi{}^u = -\frac{1}{8\pi G_N}\frac{N(\phi)}{r}  \\
        &T_u{}^\phi = O\left(\frac{1}{r^3}\right)\\
        & T_\phi{}^\phi=0.\\
    \end{split}
    \end{equation}
 Using this stress tensor, we compute the Brown-York charges associated with the BMS symmetries. This matches the ADM formula for the mass and angular momentum, as well as results from the covariant phase-space perspective \cite{Barnich:2006av,Barnich:2010eb,Prema:2021sjp}. Furthermore, we have also compared our expression for the stress tensor using the Weingarten tensor with the inverse zweinbein approach of \cite{Bagchi:2022eav,Bagchi:2021gai} (see Appendix \ref{app:free-scalar} for the comparison). 
    \item
     The stress tensor of a two-dimensional conformal Carrollian field theory transforms anomalously under the conformal Carrollian (BMS$_{3}$) algebra, with two independent central charges $c_{L}$ and $c_{M}$ appearing as the
coefficients of the inhomogeneous terms in the superrotation and
supertranslation sectors respectively. Matching this transformation law against the BMS$_{3}$ transformation of the holographic stress tensor computed above
fixes both charges in terms of bulk data,
\begin{equation}
    c_{L} = 0, \qquad c_{M} = \frac{3}{G_{N}} .
\end{equation}
The vanishing of $c_{L}$ reflects the parity invariance of pure
three-dimensional Einstein gravity, and would be lifted by a gravitational Chern-Simons term. These values agree with those obtained by an ultra-relativistic contraction of two copies of the Virasoro algebra carrying the Brown-Henneaux
central charge $c = \bar{c} = 3\ell/2G_{N}$. Under this contraction
$c_{L} = c - \bar{c}$ and $c_{M} = (c + \bar{c})/\ell$
(see Appendix~\ref{app:carroll limit}).

  \item Diffeomorphism invariance of the generating functional $W_{\rm AFS}[\lambda]$
translates into a Ward identity for the stress tensor correlators. Varying
$W_{\rm AFS}$ along a boundary diffeomorphism and using its invariance gives
\begin{equation}
    \mathcal{D}_{i}\left\langle T_{j}{}^{i}\right\rangle_{\lambda}
    + \sum_{A}\left\langle \mathcal{O}_{A}\right\rangle_{\lambda}\,
      \partial_{j}\lambda_{A}
    = \left\langle \mathcal{F}^{\rm out}_{j}\right\rangle_{\lambda},
\end{equation}
where $\mathcal{D}_{i}$ is the Carrollian connection on $\mathscr{I}^{+}$,
$\lambda_{A}$ denote the boundary sources for operators
$\mathcal{O}_{A}$, and $\mathcal{F}^{\rm out}_{j}$ is the matter flux through null
infinity. The stress tensor is therefore not conserved but obeys a flux-balance law: the failure of $\mathcal{D}_{i}\langle T_{j}{}^{i}\rangle$ to vanish measures matter flux escaping to $\mathscr{I}^{+}$, and conservation is recovered only in the absence of outgoing flux. Contracting this identity with
the vector fields generating supertranslations and superrotations yields the BMS$_{3}$ charge algebra.
    
\item The gravitational action on a null boundary is supplemented by the GHY-type term
\begin{equation}
    I_{\partial} = \frac{1}{8\pi G_{N}}\int du\, d\phi\, \sqrt{q}\,
    \left(\kappa + \Theta\right),
\end{equation}
with $\kappa$ the inaffinity of the null generators and $\Theta$ their expansion. Evaluating this term on shell for the asymptotically flat solutions above yields an action for the boundary degrees of freedom alone. Restricting to
the Minkowski sector, the on-shell boundary action reduces to a
BMS Schwarzian theory
\cite{Maldacena:2016hyu,Cotler:2018zff,Merbis:2019wgk,Cotler:2024cia},
\begin{equation}
     I_{\Theta}^{\rm mink}
    =
    \frac{1}{8\pi G_N}\int du d\phi
    \left\{\tan\frac{f(\phi)}{2},\phi\right\}
     .
\end{equation}
The boundary action has further first-order terms which give constraint equations. This is discussed in section \ref{sec:null-boundary-bms-schwarzian}.
\end{itemize}

This paper is organized as follows: In section \ref{reviewads}, we review the construction of the holographic stress tensor in AdS and asymptotically flat spacetimes. Later in section \ref{subsec: calculating stress tensor}, we explicitly compute the holographic stress tensor for AFS$_3$. In section \ref{sec: boundary theory and central charges}, we use the transformation of the holographic stress tensor under BMS$_3$ symmetries and, upon comparison with the boundary transformation rules, extract the central charges. In section \ref{sec:scalar-source-bms-ward}, we study the Ward identities satisfied by the stress-tensor and recast them in terms of flux balance laws. In section \ref{sec:null-boundary-bms-schwarzian}, we find the on-shell action for boundary graviton modes. In Appendix \ref{app:free-scalar}, we review Carrollian scalar field theory, specifically the electric, magnetic, and mixed theories, along with their stress tensors. These are classical theories, and we find their central charges to be zero as expected. In Appendix \ref{app:carroll limit}, we review the flat space limit of AdS/CFT and show how the BMS algebra (along with central extension) is derived by contracting two copies of the Virasoro algebra. More details about the connection on null infinity is discussed in appendix \ref{app:connection}. In appendix \ref{app:carroll sources}, we discuss the Carroll sources. The variation of the generating function with respect to these sources yields the one-point function of the operators. Finally, in appendix \ref{app:CFT stress tensor} we discuss the map between the Brown-York stress tensor and the boundary CFT stress tensor.

\section{The quasilocal stress tensor}
\label{reviewads}

In theories of gravity, there is no good notion of a local energy-momentum stress tensor. Usually the stress tensor consists of products of the first and zeroth derivatives of the metric, and by a coordinate transformation, one can always set such terms to zero. But in the context of holography, one instead defines a quasilocal stress tensor on the boundary of the gravitational theory. In the holographic dictionary, this stress tensor can be interpreted as the expectation value of the stress tensor of the dual quantum field theory living on the boundary.

In this section, we first review how the boundary stress tensor is defined in Anti-de Sitter spacetimes, which have a timelike boundary \cite{Balasubramanian:1999re,PhysRevD.47.1407} and then for asymptotically flat spacetimes which have a null boundary \cite{Bhambure:2024ftz,Hartong:2025jpp,Chandrasekaran:2021hxc,Fiorucci:2025twa,Chandrasekaran:2018aop}. Finally we calculate its exact form for a generic asymptotically flat spacetime in 3d.

\subsection{Stress Tensor for AdS: timelike boundary}
\label{stress tensor for AdS}

The gravitational action for $(d+1)$-dimensional AdS space with cosmological constant $\Lambda = -d(d-1)/2\ell^2$ is
 
\begin{equation}
    S = -\frac{1}{16\pi G_N}\int_{\mathcal{M}} d^{d+1}x\,\sqrt{-g}\left(R - 2\Lambda\right) - \frac{1}{8\pi G_N}\int_{\partial\mathcal{M}} d^d x\,\sqrt{-\gamma}\,K + \frac{1}{8\pi G_N}S_{\rm ct}(\gamma_{\mu\nu})
\end{equation}

The first term in the action is the Einstein-Hilbert term along with the cosmological constant term. The second term is the Gibbons–Hawking–York (GHY) boundary term, required for a well-posed variational principle \cite{Gibbons:1976ue,York:1972sj}. Some counterterms are often needed to make the conserved charge finite.  Here $\gamma_{\mu\nu}$ is the induced metric on the boundary $\partial\mathcal{M}$. Using the outward-pointing normal vector $\hat{n}^\mu$, we can find the extrinsic curvature and its trace as
 
\begin{equation}
    K_{\mu\nu} = -\tfrac{1}{2}(\nabla_\mu \hat{n}_\nu + \nabla_\nu \hat{n}_\mu), \quad K= \gamma^{\mu\nu} K_{\mu\nu}.
\end{equation}
 
 The quasilocal stress tensor (Brown-York) is obtained by varying the total action with respect to the boundary metric \cite{PhysRevD.47.1407,Balasubramanian:1999re}:
 
$$T^{\mu\nu}_{\rm ren} = \frac{2}{\sqrt{-\gamma}}\frac{\delta S}{\delta \gamma_{\mu\nu}} = \frac{1}{8\pi G_N}\left[K^{\mu\nu} - K\gamma^{\mu\nu} + \frac{2}{\sqrt{-\gamma}}\frac{\delta S_{\rm ct}}{\delta \gamma_{\mu\nu}}\right].$$
 
Without counterterms, $T^{\mu\nu}$ diverges as the hypersurface is pushed to the AdS conformal boundary. The counterterm action $S_{\rm ct}$, built from intrinsic boundary curvature invariants, cancels these divergences. Its form is fixed essentially uniquely by requiring finiteness of $T^{\mu\nu}$ and conserved charges \cite{Balasubramanian:1999re,deHaro:2000vlm,Skenderis:2002wp,Emparan:1999pm}. Via the AdS/CFT correspondence, this renormalized stress tensor is identified with the expectation value of the CFT stress tensor: $\langle T^{\mu\nu}\rangle = T^{\mu\nu}_{\rm ren}$. From now on, we drop the label `ren' as it is understood that we have a finite stress tensor.
 
%$$\text{AdS}_3{:}\quad S_{\rm ct} = -\frac{1}{\ell}\int\sqrt{-\gamma}, \qquad \text{AdS}_4{:}\quad S_{\rm ct} = -\frac{2}{\ell}\int\sqrt{-\gamma}\!\left(1-\frac{\ell^2}{4}R\right), \qquad \text{AdS}_5{:}\quad S_{\rm ct} = -\frac{3}{\ell}\int\sqrt{-\gamma}\!\left(1-\frac{\ell^2}{12}R\right).$$
 
% Unlike the original Brown–York prescription (which removes divergences by embedding the boundary in a reference spacetime), the counterterm method is always well-defined and depends only on the boundary geometry. 
 
Given a boundary Killing vector $\xi^\mu$, the associated conserved charge is
 
$$Q_\xi = \int_\Sigma d^{d-1}x\,\sqrt{\sigma}\,u^\mu T_{\mu\nu}\xi^\nu,$$
 
where $u^\mu$ is the unit normal to the spatial slice $\Sigma \subset \partial\mathcal{M}$ and $\sigma$ is the induced metric on $\Sigma$. Now, we work on an explicit example in AdS$_3$.
 
\subsubsection*{AdS$_3$ stress tensor and Brown-York charges}
 
The Poincaré patch of AdS$_3$ has the metric:
\begin{equation}
    ds^2 = \frac{\ell^2}{r^2}dr^2 + \frac{r^2}{\ell^2}(-dt^2+dx^2).
\end{equation}

Including the AdS$_3$ counterterm $S_{\rm ct} = -(1/\ell)\int\sqrt{-\gamma}$, the renormalized stress tensor for this background vanishes, $T^{\mu\nu}=0$, as expected for the global AdS (or vacuum state in CFT). For a general perturbation $\delta g_{\mu\nu}$ around this background, one finds
\begin{align}
    8\pi G_N\, T_{tt} &= \frac{r^4}{2\ell^5}\,\delta g_{rr} + \frac{\delta g_{xx}}{\ell} - \frac{r}{2\ell}\,\partial_r \delta g_{xx} \notag \\
    8\pi G_N\, T_{xx} &= \frac{\delta g_{tt}}{\ell} - \frac{r}{2\ell}\,\partial_r \delta g_{tt} - \frac{r^4}{2\ell^5}\,\delta g_{rr} \notag \\
    8\pi G_N\, T_{tx} &= \frac{1}{\ell}\,\delta g_{tx} - \frac{r}{2\ell}\,\partial_r \delta g_{tx}
\end{align}

For a rotating BTZ black hole with mass $M$ and angular momentum $J$, the metric components are $\delta g_{rr} = 8G_NM\ell^4/r^4$, $\delta g_{tt} = 8G_NM$, $\delta g_{t\phi}=-4G_NJ$. The corresponding stress tensor gives the Brown-York charges $Q_{\partial_t} = M$ and $Q_{\partial_\phi} = J$ for the appropriate Killing vector fields.
 
\subsubsection*{Extracting the central charge of dual CFT}

Asymptotic symmetries of a gravitational theory are the diffeomorphisms that preserve the asymptotic form of the metric (boundary conditions). The precise conditions for AdS$_3$ are given by Brown and Henneaux \cite{Brown:1986nw} and it transforms the metric by $\delta g_{\pm\pm} = -(\ell^2/2)\partial_\pm^3\xi^\pm$. Here, $x_{\pm}$ are the light-cone coordinates.  Computing $T_{\pm\pm}$ from this shifted metric gives
\begin{equation}
    T_{\pm\pm} = -\frac{\ell}{16\pi G_N}\,\partial_\pm^3\xi^\pm
\end{equation} 

This precisely matches the CFT$_2$ transformation law $T \to T + 2\partial\xi\,T + \xi\,\partial T - \frac{c}{24\pi}\partial^3\xi$ for the central charge
\begin{equation}
    c = \frac{3\ell}{2G_N}
\end{equation}
in agreement with the Brown-Henneaux central charge. The trace anomaly $T^\mu{}_\mu = -\frac{c}{24\pi}\mathcal{R}$ of the dual CFT$_2$ is also reproduced by evaluating the trace of the renormalized stress tensor and using the Fefferman–Graham expansion of the boundary metric \cite{FeffermanGraham1985,Henningson:1998gx}. Next, we review the analogous construction for the AFS.

\subsection{Stress Tensor for AFS (null boundary)}
\label{flatStress}
In this section, we summarize the construction of the stress tensor in asymptotically flat spacetimes (AFS). Only the parts relevant for the evaluation of the stress tensor are discussed (for details see \cite{Bhambure:2024ftz,Chandrasekaran:2021hxc,Freidel:2022vjq,Freidel:2024emv,Chandrasekaran:2018aop}).\\

Let $M$ be a three-dimensional spacetime equipped with a metric $g_{ab}$
and its Levi-Civita derivative $\nabla_a$. We consider a codimension-one
hypersurface $\mathcal{H}$ defined by $r(x) = r_0.$  The normal to the hypersurface is denoted by $n_a$. We also introduce an auxiliary vector $k^a$.
% In this foliation, the bulk metric can be written as
% \begin{eqnarray}
%     g_{ab}= q_{ab}+n_ak_b+n_bk_a
% \end{eqnarray}
% Here $n_a$ and $k_a$ are the normal and auxiliary vectors, respectively. And $q_{ab}$ is the transverse metric.
As the hypersurface is taken toward $\mathcal{I}^+$ ($r \to \infty$), its normal becomes null and the induced Carrollian data approach the corresponding data on $\mathscr{I}^{+}$ \cite{Riello:2024uvs,Freidel:2022vjq,Freidel:2024emv,Gourgoulhon:2005ng,Ciambelli:2019lap}. The hypersurface suitable for our analysis is $r-\frac{ u\, M(\phi)}{2}=\mathrm{const}$. \\

Bulk tensors are restricted to $T\mathcal{H}$ with a rigging projector
$\Pi: TM \to T\mathcal H$. The Carrollian vector can be found by projecting the normal vector onto $\mathcal{H}$ and the auxiliary and normal vectors satisfy the identities:
\begin{eqnarray}
&&{\Pi_a}^b= {\delta_a}^b + n_a k^b, \quad k^a {\Pi_{a}}^b = {\Pi_a}^b n_b = 0\nonumber\\
&&k_a k^a=0, \quad \text{while}\quad n_a n^a= 2 \Omega, \quad  \text{and} \quad \Omega \rightarrow 0 \text{ at the null infinity}\nonumber\\
&&\ell^a := n^b {\Pi_b}^a \implies \quad \ell^a= n^a + 2\Omega k^a, \quad \ell_a \ell^a= -2\Omega
\end{eqnarray}
% As an example, if $X \in T\mathcal{M}$ and $\boldsymbol{\omega} \in T^{*} \mathcal{M}$, then $\overline{X}^b := X^a {\Pi_a}^b \in T\mathcal{H}$ and $\overline{\omega}_a := {\Pi_{a}}^{b}\omega_b  \in  T^{*}\mathcal{H}$. These projected tensor satisfies $\overline{X}^a n_a = k^a \overline{\omega}_a = 0$.\\

% To treat timelike and null surfaces simultaneously, one often chooses the rigging vector to be null by requiring $\overline{k}_a:=k_a$ \cite{Freidel:2022vjq}, where $k_a=g_{ab}k^b$. Thus, we have $\overline{k}_a k^a = k_{a}k^{a} = 0$ making $k$ a \textit{null rigging vector},
% \begin{eqnarray}

% \end{eqnarray}
% The pair $(n,k)$ defines the null rigged structure on hypersurface $\mathcal{H}$. %\vaishnavi{why is some parts of n bold? And should we use i,j indices for coordinates on H}. 
% Next, the tangential vector field  $\ell= \ell^a \partial_a \in T\mathcal{H}$ can be obtained from projection of normal vector.
% \begin{eqnarray}
% \label{norm}
% \ell^a := n^b {\Pi_b}^a \implies \quad \ell^a= n^a + 2\Omega k^a, \quad \ell_a \ell^a= -2\Omega
% \end{eqnarray}
The tangential vector satisfies $\iota_{\ell} \textbf{n}=0 \text{ and} \, \iota_{\ell} \textbf{k}=-1$.  $k_a = g_{ab}k^b$ is called Ehresmann connection. The normalization $\Omega$ is of $O(1/r^{\#})$.  At the null infinity, the normal vector $n$ becomes a Carrollian vector denoted as $\ell^a$. This vector is in the kernel of the degenerate metric $q_{ab}\ell^b=0$.

% In a nutshell, the structure $(\textbf{n},k)$ at hypersurface $\mathcal{H}$ and spacetime metric $g$, we have following relations
% \begin{eqnarray}
% \ell^a = n_b g^{bc}{\Pi_c}^a, \quad k_a = g_{ab}k^b, \quad \text{and} \quad q_{ab} = {q_a}^c {q_b}^d g_{cd}  
% \end{eqnarray} 
The vectors $(\ell^a, k^a, e_A{}^a)$, together with their dual one-forms,
provide an adapted frame for the bulk geometry. In this frame, $g_{ab}$
separates naturally into transverse, Carrollian, and radial pieces:
\begin{eqnarray}
\label{withOmega}
g_{ab} &=& q_{ab}-k_a \ell_b - n_a k_b\nonumber\\
&=& q_{ab} - 2 n_{(a}k_{b)} - 2\Omega k_a k_b
\end{eqnarray}

Projecting all bulk indices with $\Pi$ defines an induced connection $D$ on $\mathcal{H}$.
For a bulk tensor $F$, the covariant derivative on $\mathcal{H}$ can be written as
\begin{eqnarray}
D_{a}{F_b}^{c}={\Pi_a}^d {\Pi_b}^e (\nabla_d {F_e}^f) {\Pi_f}^c
\end{eqnarray}
This connection is torsion-free but generally non-metric-compatible.
% The natural volume form on $\mathcal{H}$ is obtained as
% $
% \eta = \iota_k \epsilon .
% $ 
% and preserves the rigged projection tensor (see section 2.5 of \cite{Freidel:2022vjq} for more details.)
% However, the rigged metric is not covariantly constant with respect to the rigged connection. Instead, it is proportional to the extrinsic curvature of $\mathcal{H}$
% \begin{eqnarray}
% D_a H^{bc}= -({K_a}^b \ell^c+ {K_a}^c \ell^b)
% \end{eqnarray}
% where ${K_a}^b={\Pi_a}^c  \nabla_ck^d {\Pi_d}^b$ is the extrinsic curvature, sometimes called the \textit{rigged extrinsic curvature}. 
The volume form $\eta$ on $\mathcal{H}$ is obtained from spacetime volume form $\epsilon$ by doing contraction  as $\eta:= \iota_k \epsilon$. Next, we are going to study the intrinsic and extrinsic geometry at null infinity. %\vaishnavi{some parts of the above subsection are repeated in the next subsection}

\subsection*{Intrinsic and extrinsic geometry at null infinity}
On $\mathscr{I}^{+}$ we use $x^i = (u,\phi)$. The induced Carrollian data
consist of a degenerate transverse metric $q_{ij}$ and a normal vector
$n^i$, satisfying
\(n^i q_{ij} = 0.\)
We denote the transverse and full boundary volume forms by $\mu$ and $\eta$,
respectively. The volume forms and extrinsic curvature can be represented as
\begin{eqnarray}
d \mu = \Theta \eta ,  \quad K_{ij}= \frac{1}{2} \mathcal{L}_{n} q_{ij}
\end{eqnarray}
Above equation defines the expansion $\Theta$ and extrinsic curvature $K_{ij}$. Although $K_{ij}$ is fixed by $(q_{ij}, n^i)$,
it is useful to package the relevant extrinsic information in the
Weingarten map $W_i{}^j$. It is the following:
\begin{eqnarray}
{W_j}^i \equiv {\Pi_j}^a \nabla_a n^b {\Pi_b}^i, \quad {W_j}^i q_{ik} =K_{jk}, \quad {W_j}^i n^j= \kappa\, n^i.
\label{weingarten}
\end{eqnarray}
The last relation shows that the Carrollian generator
is an eigenvector of $W_i{}^j$, with eigenvalue given by the inaffinity
$\kappa$. Here we have introduced a mixed index projection tensor with bulk spacetime indices labeled by $a=1,2,3$ while boundary indices (on null infinity) are denoted by $i=1,2$. These indices are suitable for a tensor defined strictly on null infinity. 
\begin{eqnarray}
{\Pi_i}^a= {\delta_i}^a+n_i k^a, \quad {\Pi_a}^i= {\delta_a}^i+n_a k^i, \quad \text{and}\,\, {\Pi_b}^a= {\delta_b}^a+ n_b k^a= {\Pi_b}^i {\Pi_i}^a.
\end{eqnarray}
These mixed tensors can be used to find the Carrollian vector and the Ehresmann connection on null infinity as $n^i= {\Pi_a}^i n^a$  and $k_i={\Pi_i}^a k_a$, with $n^i k_i=-1$ and $k.k=0$.  The same mixed projectors induce the boundary derivative $D_i$ from the bulk Levi-Civita connection $\nabla_a$. \cite{Mars_1993} 
\begin{eqnarray}
D_i {F}^j= {\Pi_i}^a {\Pi_b}^j D_a F^b, \quad \text{where}\, \, D_a F^b= {\Pi_a}^c {\Pi_d}^b \nabla_c F^d
\end{eqnarray}
% This connection is torsionless but not metric compatible.
% as
% \begin{eqnarray}
% \label{connection}
% D_i q_{jk}=k_j K_{ik}+k_k K_{ij},\quad  D_i \eta= - \rho_i \eta\\
% D_i n^j= {W_i}^j={K_i}^j+ \rho_i n^j, \quad D_i \mu= {K_i}^j \eta_j
% \end{eqnarray}
% The connection acts on covariant vectors as %\vaishnavi{this is a little confusing}
% \begin{eqnarray}
% D_i F_j= {\Pi_i}^a {\Pi_j}^b \nabla_a( {\Pi_b}^k F_k)
% \end{eqnarray}
% The auxiliary vector $k_i$ is always 

% With this one form, one can also define a projector on the horizontal forms as
% \begin{eqnarray}
% {q_j}^i={\delta^i}_j+k_j n^i 
% \end{eqnarray}
% This also helps us in defining the inverse metric as $q_{ij}q^{jk}={q_{i}}^k.$ The shape operator can be decomposed as
% \begin{eqnarray}
% {W_j}^i={K_j}^i+\rho_j n^i , \quad \rho_j=\bar{\omega}_j- \kappa k_j
% \end{eqnarray}
% Here ${K_j}^i= q^{ik} K_{kj}$ and $\rho_j$ is called rotation one form which is defined by
% \begin{eqnarray}
% \rho_j =- {\Pi_j}^a k_b \nabla_a n^b, \quad \bar{\omega}_i= {q_i}^j \rho_j
% \end{eqnarray}
% here $\bar{\omega}$ is  H\'{a}j\'{i}\v{c}ek one form. 
\subsection*{Stress tensor}
With all this structure on the null boundaries, the stress tensor can be written as \cite{Bhambure:2024ftz,Chandrasekaran:2020wwn,Chandrasekaran:2021hxc}:
\begin{eqnarray}
    T_j{}^i= W_j{}^i-\delta_j{}^i W, \quad W= W_j{}^j
\end{eqnarray}

\subsection{Calculating the stress tensor for AFS$_3$}
\label{subsec: calculating stress tensor}
In this section, we explicitly calculate the mixed index boundary stress tensor for AFS as defined in Section \ref{flatStress}. We work in 3 spacetime dimensions with coordinates $(r,u,\phi)$, where $u$ is
retarded time, $r$ is the radial coordinate, and $\phi \sim \phi + 2\pi$ is the angular coordinate. The AFS metric is often written in the Bondi gauge. It is defined by the conditions
\begin{equation}
    g_{rr} = 0\,, \qquad g_{r\phi} = 0\,, \qquad g_{ru} = -1\,.
    \label{eq:Bondigauge}
\end{equation}
We also impose the boundary conditions required for asymptotic flatness in terms of the following fall-off \cite{Barnich:2006av}:
\begin{equation}
    g_{uu} = \mathcal{O}(r^0)\,, \qquad
    g_{u\phi} = \mathcal{O}(r^0)\,, \qquad
    g_{\phi\phi} = \mathcal{O}(r^2)\,.
    \label{eq:falloffs}
\end{equation}
The most general solution to the vacuum Einstein equations satisfying \eqref{eq:Bondigauge}
and \eqref{eq:falloffs} can be written as \cite{Barnich:2006av,Prema:2021sjp}
\begin{equation}
    ds^2 = M(\phi)\,du^2 - 2\,du\,dr
    + 2\!\left[N(\phi) + \frac{u}{2}\,\partial_\phi M(\phi)\right]du\,d\phi
    + r^2\,d\phi^2\,
    \label{eq:Bondi metric1}
\end{equation}
Here $M(\phi)$ is the Bondi mass aspect and $N(\phi)$ is the angular momentum aspect.  
The bulk manifold is denoted by $\mathcal{M}$ and $\mathscr{I}^{+}$($\mathscr{I}^{-}$) is the boundary corresponding to future (past) null infinity. The complete boundary of AFS also includes the boundary at spatial infinity $i_0$, future (past) timelike infinity $i^{\pm}$, but we will ignore these for now. For the rest of the paper, we will work with the future null infinity but all calculations are similar at the past null infinity too. Conservation of charge (incoming = outgoing) can be shown as well.

\subsection*{Normal vector:}

% We therefore tilt it into the retarded-time direction as $n_\mu dx^{\mu} = dr + \alpha(\phi)\,du$, and fix
% $\alpha$ by demanding the normal vector be null on the null infinity. 
% \begin{equation}
%     n^\mu n_\mu = g^{rr} + 2\alpha\,g^{ru} + \alpha^2 g^{uu}
%     = -M(\phi) - 2\alpha + O\!\left(\tfrac{1}{r^2}\right)=0,
% \end{equation}
% This implies $\alpha = -\tfrac{1}{2}M(\phi)$.
One can choose the radial co-normal $n_{\mu}dx^{\mu}=dr$ as the starting point. However, this is not null: using $g^{rr} = -M(\phi) +
(uM'+2N)^2/4r^2$ we get $|n| \to -M(\phi)$ at large $r$.  We therefore tilt it into the retarded-time direction as $n_\mu dx^{\mu} = dr  -\tfrac{1}{2}M(\phi)\,du$, by demanding the normal vector be null at the null infinity. We can further demand that the normal be orthogonal to some hypersurface
$\Phi(r,u,\phi)=\Phi_0$ in the bulk. This gives the normal $n_\mu=\partial_\mu\Phi$. The components can be written as
\begin{equation}
    n_\mu = \left\{1,\,-\tfrac{1}{2}M(\phi),\,-\tfrac{1}{2}uM^\prime(\phi) \right\}
    ,\quad n_{\mu}dx^{\mu}= dr - \tfrac{1}{2}M(\phi)\,du-\tfrac{1}{2}uM^\prime(\phi)d\phi,
\end{equation}
Using the inverse metric, we can raise the index as
\begin{equation}
    n^\mu 
    = \left\{\frac{1}{2} \left(\frac{N(\phi ) \left(u M'(\phi )+2 N(\phi )\right)}{r^2}-M(\phi )\right),-1,\frac{N(\phi )}{r^2}\right\} 
\end{equation}
which at the boundary becomes:
\begin{eqnarray}
    n^\mu  = \{ -\tfrac{1}{2}M(\phi),-1,0\}, \quad  n^\mu n_\mu = \frac{N(\phi )^2}{r^2}
\end{eqnarray}
The norm
vanishes on the boundary and is spacelike in the bulk. The normal vector satisfies $dn=-\frac{1}{2}M^\prime(\phi)d\phi\wedge du-\frac{1}{2}M^\prime(\phi)du\wedge d\phi=0$. Hence $n_\mu$ defines a bulk foliation corresponding to $r-\tfrac{1}{2}uM(\phi)=\text{constant}$ surfaces which are timelike in the bulk. For pure Minkowski ($M=-1$, $N=0$) this gives $r + \tfrac{1}{2}u=\text{const}$ and $n^\mu n_\mu = 0$ everywhere, so the foliation is null too.

Next, we choose the auxiliary null vector $k^\mu$ as:
\begin{equation}
    k^{\mu}=\{-1,0,0\}
    \label{eq: auxiliary vector}
\end{equation}
which gives $k_\mu = \{ 0,1,0\}$ and it is null $k^{\mu}k_{\mu}=0$ everywhere in the bulk. The role played by an orthogonal foliation in the timelike boundary is played here
by the auxiliary rigging vector $k^\mu = -\partial_r$, which is transverse to $\mathscr{I}^+$ and normalized so that $k^\mu n_\mu = -1$. These satisfy the following decomposition of the bulk metric
\begin{eqnarray}
\nonumber g_{ab} = q_{ab} - 2 n_{(a}k_{b)} - 2\Omega k_a k_b
\end{eqnarray}

\subsection*{Stress Tensor}
With these normal and auxiliary vectors, we can explicitly compute the stress tensor. First, we evaluate the Weingarten tensor $W_j{}^i$ and then the stress tensor $T_j{}^i$. The components of the stress tensor are (in $u,\phi$ coordinates):
% \begin{equation}
% \label{stresstensor}
%     T_j{}^i = -\frac{1}{8 \pi G_N}\left(
% \begin{array}{cc}
%  \frac{M(\phi )}{2 r}+\frac{-\frac{1}{2} u M''(\phi )-N'(\phi )}{r^2} & \frac{\frac{1}{2} u M'(\phi )+N(\phi )}{r} \\
%  \frac{M'(\phi )}{2 r^2}-\frac{M(\phi ) \left(u M'(\phi )+2 N(\phi )\right)}{4 r^3} & 0 \\
% \end{array}
% \right)
% \end{equation}
\begin{eqnarray}
\label{stresstensor}
  && T_u{}^u= -\frac{1}{8 \pi G_N}\left[ \frac{M(\phi )}{2 r}+O\left(\frac{1}{r^2}\right)\right] \\
  && T_{\phi}{}^u= -\frac{1}{8 \pi G_N}\left[\frac{N(\phi )}{r}\right]\nonumber\\
  && T_{u}{}^{\phi}= O\left(\frac{1}{r^3}\right)\nonumber\\
   && T_{\phi}{}^{\phi}=0
\end{eqnarray}

% \begin{eqnarray}
% \label{stresstensor}
%   && T_u{}^u= -\frac{1}{8 \pi G_N}\left( \frac{M(\phi )}{2 r}-\frac{N'(\phi )}{r^2}+\frac{u N(\phi)M^\prime(\phi)}{2r^3}\right)\nonumber\\
%   && T_{\phi}{}^u= -\frac{1}{8 \pi G_N}\left(\frac{N(\phi )}{r}\right)\nonumber\\
%   && T_{u}{}^{\phi}= -\frac{1}{8 \pi G_N}\left(-\frac{M(\phi )  N(\phi)}{2 r^3}\right)\nonumber\\
%    && T_{\phi}{}^{\phi}=0
% \end{eqnarray}

For the case of pure Minkowski, this simplifies to: %\vaishnavi{make it components}
\begin{equation}
T_u{}^u=\frac{1}{8 \pi G_N} \frac{1}{2r} \;, \qquad T_\phi{}^u=T_u{}^\phi=T_\phi{}^\phi=0
\end{equation}

\subsection*{Brown York charges from the stress tensor}
Once we have the mixed index stress tensor, the Brown-York charges can be calculated \cite{Bhambure:2024ftz,Chandrasekaran:2021hxc} by contracting the stress tensor with the Killing vector field $\xi$ and auxiliary vector (see also \cite{Ashtekar:1981bq,Wald:1999wa} for the Wald-Zoupas charges)
\begin{eqnarray}
    Q = -\int_S T_j{}^{\,i}\,  \xi^j\, k_i \, \mu.
\end{eqnarray}
Here $\xi^i$  generates the BMS symmetry transformations which is 
\begin{equation}
    \xi = \big(T(\phi) + u\,\partial_{\phi} R(\phi)\big)\partial_u
          + R(\phi)\,\partial_{\phi} ,
\end{equation}
composed of two arbitrary functions on the circle: $T(\phi)$ are called \emph{supertranslations} and $R(\phi)$ are \emph{superrotations}.

$k_i$ is the auxiliary co-vector $k_i=\{0,1,0\}$ introduced above, and $\mu= r d\phi $ is the volume element on a spatial cut $S$ of $\mathscr{I}^+$. Using \eqref{stresstensor} this becomes:
\begin{eqnarray}
    Q = -\int_S \mu(T_u{}^u \xi^u + T_\phi{}^u \xi^\phi)
\end{eqnarray}
Furthermore, we can use the Fourier mode expansion of $\xi$,
\begin{equation}
\begin{split}
    P_m & = \xi(T=e^{im\phi},R=0) = e^{im\phi} \partial_u \\ 
    J_m & = \xi(T=0,R=e^{im\phi}) = i m u e^{im\phi}\partial_u + e^{im\phi}\partial_\phi \\
\end{split}
\end{equation}
to calculate their corresponding charges as:
\begin{equation}
    \begin{split}
        Q(P_m)\equiv \mathcal{P}_m & = -\int_S \mu T_u{}^u e^{im\phi} \\
        & = \frac{1}{16 \pi G_N} \int_0^{2\pi} d\phi M(\phi) e^{im\phi} \\
        Q (J_m)\equiv\mathcal{J}_m & = -\int_S \mu(T_u{}^u\, imue^{im\phi}+ T_\phi{}^u e^{im\phi}) \\
        & = \frac{1}{8\pi G_N}\int_0^{2\pi}d\phi \left( \frac{M(\phi)}{2} i m u e^{im\phi} + N(\phi) e^{im\phi} \right)  \\
        & = \frac{u}{16\pi G_N} \left( M(2\pi)e^{i2\pi m} - M(0) -\int_0^{2\pi}d\phi M^\prime(\phi)e^{im\phi} \right) + \frac{1}{8\pi G_N}\int_0^{2\pi} d\phi N(\phi) e^{im\phi} \\
        & = \frac{1}{8\pi G_N}\int_0^{2\pi} d\phi \left( N(\phi)-\frac{1}{2}uM^\prime(\phi) \right) e^{im\phi}
    \end{split}
\end{equation}
%\vaishnavi{I think we have a problem here. With the new normal the M' terms will disappear so it wont cancel with the M integral. So the final angular momentum charge will have an integral over (N-M')exp. I have made the changes in normal till the previous section, lets talk tomorrow and then I will change the rest}
%\hare{don't worry about it, change the rest. }

In the second last line we have used that \(m\) is integer and periodic \(M(\phi)\),
\begin{equation}
\left[
M(\phi)e^{im\phi}
\right]_{0}^{2\pi}
=
M(2\pi)e^{2\pi im}-M(0)=0.
\end{equation}
Furthermore, we can do the mode decomposition
\begin{equation}
M(\phi)=\sum_{k\in \mathbb{Z}} M_k e^{-ik\phi},
\qquad
N(\phi)=\sum_{k\in \mathbb{Z}} N_k e^{-ik\phi},
\end{equation}
with
\begin{equation}
M_k=\frac{1}{2\pi}\int_0^{2\pi} d\phi\, M(\phi)e^{ik\phi},
\qquad
N_k=\frac{1}{2\pi}\int_0^{2\pi} d\phi\, N(\phi)e^{ik\phi}.
\end{equation}
Therefore
\begin{equation}
\mathcal{P}_m
=
\frac{1}{16\pi G_N}
\int_0^{2\pi}d\phi\,M(\phi)e^{im\phi}
=
\frac{M_m}{8G_N}.
\end{equation}
Similarly,
% \begin{equation}
% \begin{split}
% \mathcal{J}_m
% &=
% \frac{1}{8\pi G_N}
% \int_0^{2\pi}d\phi
% \left[
% \frac{u}{2}M'(\phi)e^{im\phi}
% +
% \frac{imu}{2}M(\phi)e^{im\phi}
% +
% N(\phi)e^{im\phi}
% \right] \\
% &=
% \frac{1}{8\pi G_N}
% \int_0^{2\pi}d\phi
% \left[
% \frac{u}{2}
% \left(M'(\phi)+imM(\phi)\right)e^{im\phi}
% +
% N(\phi)e^{im\phi}
% \right] \\
% &=
% \frac{1}{8\pi G_N}
% \left\{
% \frac{u}{2}
% \int_0^{2\pi}d\phi\,
% \partial_\phi\left(M(\phi)e^{im\phi}\right)
% +
% \int_0^{2\pi}d\phi\,N(\phi)e^{im\phi}
% \right\} \\
% &=
% \frac{1}{8\pi G_N}
% \left\{
% \frac{u}{2}
% \left[
% M(\phi)e^{im\phi}
% \right]_{0}^{2\pi}
% +
% \int_0^{2\pi}d\phi\,N(\phi)e^{im\phi}
% \right\}.
% \end{split}
% \end{equation}

\begin{equation}
\mathcal{J}_m
=
\frac{1}{8\pi G_N}
\int_0^{2\pi}d\phi\left( N(\phi) + i m u \frac{M(\phi)}{2} \right) e^{im\phi}
=
\frac{N_m}{4G_N} +i m u \frac{M_m}{8 G_N}.
\end{equation}
Thus
\begin{equation}
\mathcal{P}_0=\frac{M_0}{8G_N},
\qquad
\mathcal{J}_0=\frac{N_0}{4G_N}.
\end{equation}
% We can make the shift $M\rightarrow M+1$ in $\mathcal{P}_m$ to match the normalization for Minkowski. We can see these charges calculated using the stress tensor exactly match the Noether charges found from the BMS$_3$ algebra in \eqref{eq:Pcharge} and \eqref{eq:Jcharge} (see also \cite{Bagchi:2012xr}).\vaishnavi{need to replace the eqref here with correct ones}
Total Bondi mass and angular momentum are the zero modes of these charges.

\subsection*{Stress Tensor transformation}

The BMS$_3$ transformation preserves the gauge choice and boundary conditions. These transformations act on the coordinates as 
\begin{equation}
    \begin{split}
   \delta u & = T + u \partial_\phi R \\     
   \delta r & = -r \partial_\phi R + \partial_\phi^2 T + u \partial_\phi^3 R - \frac{1}{r} (\partial_\phi T + u \partial_\phi ^2 R)(N+ \frac{u}{2} \partial_\phi M) \\
   \delta\phi & = R - \frac{1}{r} \partial_\phi T - \frac{u}{r} \partial_\phi^2 R.
    \end{split}
\end{equation}
% BMS transformations are parametrized by the 2 functions $T(\phi)$ and $R(\phi)$, which generate supertranslations and superrotations, respectively. 
As these transformations act on the coordinates, they also transform the Bondi mass and angular momentum aspects as follows:
\begin{equation}
\begin{split}
       \delta M &= R \partial_\phi M + 2 M \partial_\phi R - 2\partial_\phi^3 R   \\
       \delta N &= R \partial_\phi N + 2 N \partial_\phi R + \frac{1}{2} T \partial_\phi M + M \partial_\phi T - \partial_\phi^3 T \\
\end{split}
\end{equation}

The transformed metric retains the form \eqref{eq:Bondi metric1} but with the new functions $M\rightarrow M+\delta M$ and $N \rightarrow N+\delta N$. The metric can be expanded as
\begin{equation}
\begin{split}
     ds^2 &= d\phi^2 r^2-2 dr du+du^2 M(\phi )+d\phi du \left(u M'(\phi )+2 N(\phi )\right)\\
    & + du^2 \left(R(\phi ) M'(\phi )+2 M(\phi ) R'(\phi )-2 R^{(3)}(\phi )\right) \\
    & + d\phi  du \Big(u R(\phi ) M''(\phi )+3 u M'(\phi ) R'(\phi )+T(\phi ) M'(\phi )+2 u M(\phi ) R''(\phi )\\
    &+2 M(\phi ) T'(\phi )-2 u R^{(4)}(\phi ) +4 N(\phi ) R'(\phi )+2 R(\phi ) N'(\phi )-2 T^{(3)}(\phi )\Big) \\
\end{split}
\end{equation}
The normal 1-form is similarly transformed to
\begin{equation}
    n_{\mu}=\left\{1,-\frac{1}{2} \left( M+\delta M\right),-\frac{1}{2}u(M^\prime+\delta M^\prime)\right\}
\end{equation}
while the auxiliary vector is still $k^{\mu}=\{-1,0,0\}$ with normalization as $n.k=-1,\; k.k=0$.
We consider an infinitesimal BMS transformation and keep only the terms linear in the transformation parameters $R(\phi)$ and $T(\phi)$, and the leading-order terms in $1/r$. This gives the following transformed boundary stress tensor: 
\begin{eqnarray}
    &&T_u{}^u=-\frac{1}{8\pi G_N}\frac{1}{r}\Big(\frac {M}{2 }+\left[\frac12 R M'+M R'-R^{(3)}\right] + O\left(\frac{1}{r^2}\right)\nonumber \\
    && T_{\phi}{}^u= -\frac{1}{8\pi G_N}\frac{1}{r}\left(N +\left[\frac{1}{2}TM^\prime + 2NR^\prime+MT^\prime+RN^\prime-T^{(3)} \right] \right)\nonumber\\
&&T_u{}^{\phi}= O\left(\frac{1}{r^3}\right)\nonumber\\
&&T_{\phi}{}^{\phi}=0
\end{eqnarray}

% \begin{equation}
% T_j{}^i=-\frac{1}{8\pi G_N}
% \left(
% \begin{array}{cc}
% \frac12 M+\epsilon\left(\frac12 R M'+M R'-R^{(3)}\right)
% &
% \begin{aligned}
% &\frac12 u M'+N
% +\epsilon\Big(
% \frac12 u R M''
% +\frac12 M'(3 u R'+T)  \\
% &\qquad
% +u M R''+M T'
% -u R^{(4)}
% +2N R'
% +R N'
% -T^{(3)}
% \Big)
% \end{aligned}
% \\[1.2em]
% O\left(\frac{1}{r}\right) & 0
% \end{array}
% \right)
% \end{equation}

\textbf{For pure Minkowski}: We set $M=-1$ and $N=0$ which gives
\begin{equation}
\begin{split}
    T_u{}^u & = \frac{1}{8\pi G_N\, r} \left( \frac{1}{2}+ \left(R'''+R'\right) \right) \\
    T_\phi{}^u & = \frac{1}{8\pi G_N\, r}   \left( T'''+T'\right) \\
    T_u{}^\phi & = O\left(1/r^3\right) \\
    T_\phi{}^\phi & = 0
\end{split}
\label{transformedbdystresstensor for Minkowski}
\end{equation}

%\vaishnavi{made changes till here so far. have to do section 3 now}
%\vaishnavi{write above in components instead}
This is the holographic stress tensor after the spacetime is transformed by supertranslations and superrotations. The CFT stress tensor can be identified by stripping a factor of $1/r$ from the holographic computation (see appendix \ref{app:CFT stress tensor}). This is the exact Carrollian counterpart of the definition of the
holographic stress tensor in asymptotically AdS spacetimes
\cite{Balasubramanian:1999re,deHaro:2000vlm}. 
\begin{equation}
    T_j{}^{i}\big|_{(R)}
    =
    \frac{1}{R}\,\mathcal T_j{}^{i}
    +
    \mathcal O\!\left(R^{-2}\right).
    \label{eq:radial-expansion-of-T}
\end{equation}
So the boundary stress tensor is
\begin{equation}
    \mathcal T_j{}^{i}
    \equiv
    \lim_{R\to\infty}R\;T_j{}^{i}\big|_{(R)},
    \qquad
    \delta W
    =
    \int_{\mathscr I^+}\eta^{(0)}\,
    \mathcal T_j{}^{i}\,\delta \mathcal C_i{}^j
    +\ldots
    \label{eq:boundary-stress-tensor-definition}
\end{equation}
$\delta \mathcal C_i{}^j$ represents the variation of the sources. %\vaishnavi{should it be $\delta C_i^j$ in both cases}
The idea is rather simple. The volume form on a hypersurface $r=R$ is given by $\eta= R du d\phi$. The variation of the action $\delta S= \int \eta T_{i}^j \delta \mathcal{C}_j^i$ shouldn't depend on radial location. For the null infinity $\eta^{(0)}= du d\phi$, hence the boundary stress tensor should be defined by stripping the factor of $1/r$. In the subsequent section, when we talk about the stress tensor, we always mean the boundary stress tensor and, by abuse of notation, still write it as $T_j{}^i$. 

\section{Boundary theory and its central charges}
\label{sec: boundary theory and central charges}

In the previous section, we studied the holographic stress tensor for AFS and its transformation under BMS symmetry. Next, we study stress tensor for the dual boundary theory which is a Carrollian/BMS CFT. We study how its stress tensor transforms under BMS transformations. By comparing these transformations to the bulk gravity theory, we identify the central charges of the dual theories similar to the exercise done in AdS case (see section \ref{stress tensor for AdS}). 
% In this section we review the basics of Carrollian structure arising from the $c\rightarrow0$ limit, the contraction of conformal symmetry, the Carrollian stress tensor and the anomalous transformation of its components from which the central charges are read off.

% \subsection{Carrollian CFT$_2$ : review}
 
% The conjectured holographic dual of three-dimensional asymptotically flat gravity is a
% field theory carrying exactly this structure intrinsically: a two-dimensional
% \emph{Carrollian conformal field theory} (CCFT$_2$), equivalently a BMS$_3$-invariant field
% theory (BMSFT$_2$).
% The asymptotic symmetry analysis of flat spacetimes (see Appendix \ref{} for a review) attaches the infinite dimensional symmetry algebra (the BMS$_3$ algebra \eqref{eq:BMSalgebra}--\eqref{eq:centralcharges}) to 

\subsection*{The Carrollian limit and Carrollian structure}
The boundary at $\mathscr{I}^+$ is degenerate and has a Carrollian structure.
It is a pair
$(q_{ab},\,\ell^a)$ where $q_{ab}$ is a positive, rank-$(d-1)$ degenerate ``transverse metric''
 and $\ell^a$ is a nowhere-vanishing vector field  in its kernel,
\begin{equation}
    q_{ab}\,\ell^b = 0 .
    \label{eq:carrkernel}
\end{equation}
This data is accompanied by an Ehresmann connection (clock one-form) $k_a$ normalized so
that $k_a\ell^a = -1$. These split the tangent bundle into the Carrollian time direction
$\ell^a$ and the transverse space. \footnote{
A Carrollian theory can also be found by  $c\to0$ (ultra-relativistic) contraction of a relativistic
theory. In this limit, the light cones close up on the time axis. In $d$ spacetime
dimensions, one rescales the boost and Hamiltonian generators of the Poincaré algebra and
sends $c\to 0$, obtaining the Carroll algebra.  The boosts commute and the time
translation $H$ becomes central \cite{Duval:2014uoa,Ruzziconi:2026bix}. The physical feature of this theory is ultralocality: all distinct spatial points are space-like separated and decouple (if the Carrollian theory is local; see \cite{Cotler:2025npu} for recent work on non-local Carrollian CFTs).}

The dual theory which lives at $\mathscr{I}^+$ have the coordinates $(u,\phi)$. Here  $u$ runs along the degenerate time direction and $\phi$ is the spatial circle with $\phi\sim\phi+2\pi$.
% In the bulk we can identify these coordinates with the Bondi coordinates,
% $(\tau,\sigma)\equiv(u,\phi)$. 
The flat Carrollian background is
\begin{equation}
    q_{ab}\,dx^a dx^b = d\phi^2 ,
    \qquad
    \ell^a\partial_a = \partial_u ,
    \qquad
    k_a\,dx^a = -\,du ,
    \label{eq:flatcarr}
\end{equation}
Starting with Bondi metric
\begin{equation}
    ds^2 = M(\phi)du^2  -2 dudr +2 \Big[N(\phi) + \frac{u}{2} \partial_\phi M(\phi) \Big]dud\phi + r^2 d\phi^2 ,
\end{equation}
one can obtain the metric and Carroll structure at null infinity by first taking $dr=0$ and pulling up a Weyl factor $r^2$.
\begin{equation}
    ds^2 =r^2\Bigg( M(\phi)\frac{du^2}{r^2}   +\frac{2}{r^2} \Big[N(\phi) + \frac{u}{2} \partial_\phi M(\phi) \Big]dud\phi +  d\phi^2 \Bigg).
\end{equation}
By taking the limit $r\rightarrow\infty$, the (conformal) metric at $\mathscr{I}^+$ can be written as $ds^2= 0+d\phi^2$. It has a Carrollian structure, and the Carrollian vector is $\ell^a\partial_a = \partial_u$.

\subsection*{Carrollian Conformal symmetry and the BMS$_3$ algebra}

The symmetries of the flat Carrollian background \eqref{eq:flatcarr} are the diffeomorphisms 
$\xi$ that preserve $(q_{ab},\ell^a)$ up to a Carrollian Weyl rescaling,
\begin{equation}
    \mathcal{L}_{\xi}\, q_{ab} = 2\,\alpha\,q_{ab},
    \qquad
    \mathcal{L}_{\xi}\, \ell^a = -\,\alpha\,\ell^a .
    \label{eq:confkilling}
\end{equation}
In two dimensions, the general solution is
\begin{equation}
    \xi = \big(T(\phi) + u\,\partial_{\phi} R(\phi)\big)\partial_u
          + R(\phi)\,\partial_{\phi} ,
    \qquad
    \alpha = \partial_{\phi} R(\phi),
    \label{eq:CCvector}
\end{equation}
generated by two arbitrary periodic functions: $T(\phi)$ are \emph{supertranslations}
(angle-dependent shifts of the Carrollian time) and $R(\phi)$ are \emph{superrotations}
(conformal diffeomorphisms of the circle).
% This coincides, at $r\to\infty$, with the
% restriction to $\mathcal{I}^+$ of the bulk Killing vector \eqref{eq:BMSvecfields} and induces
% the boundary transformations \eqref{eq:bdytransform} as expected.
Upon these symmetries,
$(u,\phi)$ transform as
\begin{equation}
    \phi \;\to\; \phi + \epsilon\,R(\phi)\,,
    \qquad
    u \;\to\; u + \epsilon\,T(\phi) + \epsilon\,u\,\partial_\phi R(\phi)\,.
    \label{eq:bdytransform}
\end{equation}
These functions can be expanded on the cylinder and the corresponding modes for the generator ($\ell_n$ generate
superrotations and $m_n$ generate supertranslations) can be written as
\begin{equation}
    \ell_n = i\,e^{in\phi}\big(\partial_\phi + i n\,u\,\partial_u\big),
    \qquad
    m_n = i\,e^{i n\phi}\,\partial_u ,
    \qquad n\in\mathbb{Z}.
    \label{eq:CCgenerators}
\end{equation}
These modes close under the Lie bracket into the (centerless) conformal Carrollian algebra in two
dimensions,
\begin{equation}
    [\ell_m,\ell_n] = (m-n)\,\ell_{m+n},
    \qquad
    [\ell_m,m_n] = (m-n)\,m_{m+n},
    \qquad
    [m_m,m_n] = 0 .
    \label{eq:CCalgebra}
\end{equation}
This is the classical BMS$_3$ algebra, which is isomorphic to the conformal Carrollian algebra
$\mathfrak{bms}_3 \simeq \mathfrak{ccarr}_2$. And in $d=2$, it is also isomorphic to the two-dimensional Galilean conformal algebra \cite{Bagchi:2009pe, Ruzziconi:2026bix,Barnich:2006av,Barnich:2012rz}. For our AFS spacetime (Einstein gravity), the asymptotic symmetries are BMS$_3$ which has a central extension.
\begin{equation}
\begin{aligned}
    [L_m,L_n] &= (m-n)\,L_{m+n} + \frac{c_L}{12}\,m(m^2-1)\,\delta_{m+n,0}, \\[2pt]
    [L_m,M_n] &= (m-n)\,M_{m+n} + \frac{c_M}{12}\,m(m^2-1)\,\delta_{m+n,0}, \\[2pt]
    [M_m,M_n] &= 0 ,
\end{aligned}
    \label{eq:CCcentral}
\end{equation}
The central charges are $c_L=0, c_M=3/G_N$ \cite{Barnich:2006av}. Note that the generators of the centerless BMS algebra are denoted by $\ell_n,m_n$ while the generators which close with a central extension are denoted by $L_n,M_n$.

\subsection*{The Carrollian stress tensor}
\label{sec:carrstress}

As in the relativistic case, the stress tensor of a theory is obtained by varying its action with respect to the background. Varying the action with Carrollian background $(\ell^a, q^{ab})$ defines a mixed-index Carrollian
stress tensor $T_{b}{}^a$ \cite{Ruzziconi:2026bix,Campoleoni:2022wmf,Hartong:2025jpp,Bagchi:2024gnn} (see also appendix \ref{app:stress} for stress tensor for electric/magnetic and mixed theories). Invariance of the action under
boundary diffeomorphisms, Carrollian boosts, and Weyl rescalings impose more constraints on the stress tensor. On the flat
background \eqref{eq:flatcarr} these become:
\begin{equation}
    \partial_a T_{b}{}^a = 0 ,
    \qquad\quad
    T_{u}{}^{\phi} = 0 ,
    \qquad\quad
    T_{a}{}^a =  T_{u}{}^{u} +  T_{\phi}{}^{\phi} = 0 .
    \label{eq:carrconditions}
\end{equation}
The first equation is the conservation of the stress tensor. The second is the Carrollian boost Ward identity, which encodes the ultralocal nature of the field theory. The third condition imposes the tracelessness of the stress tensor, required by (classical) conformal Carrollian invariance.  These conditions leave two independent components $ T_{u}{}^{u}$ which is the energy density and $T_{\phi}{}^{u}(u,\phi)$ which is momentum density. \footnote{In the notation of \cite{Hao:2021urq} these components
are written $M\equiv T_{u}{}^{u}$ and $T\equiv T_{\phi}{}^u$} 

This is the structure realized holographically. The mixed-index boundary stress tensor
computed in Section~\ref{flatStress} has $T_{u}{}^{\phi}\to 0$
 and $T_{\phi}{}^{\phi}=0$, with the
two non-trivial components $T_{u}{}^{u}$ and $T_{\phi}{}^{u}$. For AFS, the bulk trace $T^{a}_{a}\propto M(\phi)$ does \emph{not} vanish. But we do not understand its connection with the Carrollian Weyl anomaly \cite{Bagchi:2021gai} yet.

\subsection*{Transformation of the stress tensor under BMS}
\label{sec:carrtransf}

A finite BMS$_3$ transformation acts on the spatial coordinate $\phi$ and null coordinate $u$ as %\vaishnavi{convert all x, y to $\phi$ and u}
\begin{equation}
    \phi \;\to\; \tilde{\phi} = f(\phi),
    \qquad
    u \;\to\; \tilde{u} = f'(\phi)\,u + g(\phi),
    \label{eq:finite BMS}
\end{equation}
where $f$ encodes a superrotation and $g$ a supertranslation. %These transformations can also be used to map a plane to a cylinder and vice versa \eqref{eq: cylinder to plane map}. 
Let $T_1$ and $T_2$ be the Noether currents of translation along $\phi$ and $u$. %\footnote{$T_1$ and $T_2$ has been used for translations along $x$ and $y$ in \cite{Hao:2021urq}}. 
Under \eqref{eq:finite BMS} the two components of the stress tensor transform with anomalous,
inhomogeneous pieces controlled by the central charges  \cite{Bagchi:2025vri,Barnich:2010eb,Barnich:2006av}: 
\begin{equation}
\begin{aligned}
    \widetilde{T_2}(\phi)
      &= f'^{\,2}\,T_2(\tilde{\phi})
         + \frac{c_M}{24\pi}\,\{f,\phi\}, \\[4pt]
    \widetilde{T_1}(\phi,u)
      &= f'^{\,2}\,T_1(\tilde{\phi},\tilde{u})
         + 2 f'g'T_2(\tilde{\phi})+(f^\prime)^2 g T_2^\prime(\tilde{\phi})
         + \frac{c_L}{24\pi}\,\{f,\phi\}
         + \frac{c_M}{24\pi} \mathcal{S}_{\rm BMS}\left[ (f,g),\phi \right],
\end{aligned}
\label{eq:stress transf}
\end{equation}
where $\{f,\phi\}$ is the ordinary Schwarzian derivative and the last term is the ``BMS (Carrollian) Schwarzian''. These are defined as:
\begin{equation}
    \{f,\phi\} = \frac{f'''}{f'} - \frac{3}{2}\left(\frac{f''}{f'}\right)^2 ,
    \qquad
    \mathcal S_{\rm BMS}[(f,g),\phi] = \frac{3 (f'')^2-f' f''' g' - 3 f' f'' g'' + (f')^2 g'' }{(f')^3}.
    \label{eq:schwarzians}
\end{equation}
 The homogeneous pieces $f'^{\,2}T_2$
and $f'^{\,2}T_1$ are the ``tensorial'' parts and the
Schwarzian/BMS-Schwarzian terms are the anomalies whose coefficients are $c_M$ and
$c_L$.\footnote{Several conventions for \eqref{eq:stress transf} appear in the literature, differing
by the redefinition $T \rightarrow T-u\,\partial_\phi M$ noted in footnote~1
of \cite{Hao:2021urq}. We use the convention of \cite{Bagchi:2025vri}.} We compare these to the transformed gravity stress tensor and obtain the anomaly $c_M$ and $c_L$.

The future null infinity $\mathscr{I}^+$ is also a null cylinder with spatial coordinate $\phi$ and the null coordinate $u$. The infinitesimal BMS transformations acts as
\begin{equation}
    \phi
    \longrightarrow
    \phi+\varepsilon\,R(\phi),
    \qquad
    u
    \longrightarrow
    u+\varepsilon
    \left[
        u\,\partial_\phi R(\phi)
        +
        T(\phi)
    \right],
    \label{eq:inf-bms-cylinder}
\end{equation}
 This can be written as \eqref{eq:finite BMS} with:
\begin{equation}
    f(\phi)=\phi+\varepsilon\,R(\phi),
    \qquad
    g(\phi)=\varepsilon\,T(\phi).
    \label{eq:inf-bms-fg}
\end{equation}

The two BMS currents \(T_1\) and \(T_2\) can be identified with the boundary stress-tensor components as
\begin{equation}
    T_2 \equiv T_{u}{}^u,
    \qquad
    T_1 \equiv T_{\phi}{}^u.
    \label{eq:T1-T2-identification}
\end{equation}

For the Minkowski vacuum background, the relevant boundary stress tensor is
\begin{equation}
    T_2(\phi)=\frac{1}{16\pi G_N},
    \qquad
    T_1(\phi,u)=0.
    \label{eq:minkowski-bdy-stress}
\end{equation}

We calculate the Schwarzian and BMS Schwarzian to
linear order in \(\varepsilon\).
\begin{equation}
    \{f,\phi\}
    =
    \varepsilon\,R'''(\phi)
    +
    \mathcal O(\varepsilon^2),
    \label{eq:schwarzian-linear}
\end{equation}
and
\begin{equation}
    \mathcal S_{\rm BMS}[(f,g),\phi]
    =
    \varepsilon
        T'''(\phi)
    +
    \mathcal O(\varepsilon^2).
    \label{eq:bms-schwarzian-linear}
\end{equation}
Using the vacuum values \eqref{eq:minkowski-bdy-stress}, the transformed
currents are therefore \footnote{The currents used in \cite{Bagchi:2025vri} are normalized
as CFT currents.  To compare them with the gravitational boundary
stress-tensor densities, we define
\(T_1=(2\pi)^{-1}T_1^{CFT}\) and
\(T_2=(2\pi)^{-1}T_2^{CFT}\).
This rescaling leaves the homogeneous terms in the transformation laws
unchanged but converts the anomalous coefficients
\(c_{L,M}/12\) into \(c_{L,M}/(24\pi)\).
The central charges themselves are not rescaled.} %\vaishnavi{I have modified this comment check if it still makes sense, otherwise remove it} \hare{make sense}}
\begin{equation}
    \widetilde{T}_2(\phi)
    =
    \frac{1}{16\pi G_N}
    +
    \varepsilon
    \left[
        \frac{ c_M}{24 \pi}R'''(\phi)
        +
        \frac{1}{8\pi G_N}R'(\phi)
    \right]
    +
    \mathcal O(\varepsilon^2),
    \label{eq:T2-inf-transform}
\end{equation}
and
\begin{equation}
    \widetilde{T}_1(\phi,u) =
    \varepsilon
    \left[ \frac{1}{8\pi G_N} T'(\phi) + \frac{ c_L}{24\pi} R'''(\phi) + \frac{ c_M}{24 \pi} T'''(\phi) \right] + \mathcal O(\varepsilon^2).
    \label{eq:T1-inf-transform}
\end{equation}

Comparing \eqref{eq:T2-inf-transform} and \eqref{eq:T1-inf-transform} with
the direct gravitational transformation of the boundary stress tensor \eqref{transformedbdystresstensor for Minkowski} fixes the central charges to be

\begin{equation}
    c_L=0,
    \qquad
    c_M=\frac{3}{G_N}.
    \label{eq:bms3-central-charges}
\end{equation}
These are the standard central charges of three-dimensional Einstein gravity in the flat-space limit.  Although the above comparison was carried out for the Minkowski background for simplicity, the same values are obtained by
starting from the general Bondi metric and comparing the transformation of the boundary stress tensor.\\

\section{Ward identities from Carrollian stress tensor}
\label{sec:scalar-source-bms-ward}

In the previous sections, we evaluated the stress tensor for asymptotically flat spacetime. Its conservation leads to the Ward identity. We now derive the Ward identities obeyed by correlators of the boundary Carrollian stress tensor (see also \cite{Fiorucci:2025twa,Bagchi:2015wna,Bagchi:2016geg}) with scalar operator insertions (see recent discussion on Ward identities by \cite{Hartong:2025jpp,Hartong:2026rbr}).

\paragraph{Generating functional and one-point functions:}
Future null infinity is the two-dimensional Carrollian manifold
\(\mathscr I^+\simeq\mathbb R_u\times S^1_\phi\). The Carroll data is
\begin{equation}
    \bigl(q_{ij},\,n^i,\,k_i\bigr),
    \qquad
    q_{ij}n^j=0,
    \qquad
    k_in^i=-1,
    \label{eq:carroll-data}
\end{equation}
with \(q_{ij}\) being the degenerate spatial metric, \(n^i\) the null generator, and \(k_i\) the Ehresmann connection. 

% The normalisation
% \(k_in^i=-1\) is the standard Carrollian one, and it makes \(n^i\)
% past-directed. One can introduce the future-directed generator
% \begin{equation}
%     \ell^i\equiv-n^i,
%     \qquad
%     k_i\ell^i=+1,
%     \qquad
%     q_{ij}\ell^j=0,
%     \label{eq:future-generator}
% \end{equation} 

We define the
renormalized generating functional
\begin{equation}
    W_{\rm AFS}[\mathcal C,\lambda]
    =
    -\,I_{\rm ren}^{\rm on\text{-}shell}[\mathcal C,\lambda],
    \label{eq:afs-generating-functional}
\end{equation}
where \(\mathcal C\) denotes the Carrollian boundary data and \(\lambda(u,\phi)\) is the source for a scalar operator
\(\mathcal O(u,\phi)\). The variation of sources i.e. \(\delta \mathcal C\) can be established as (see full details in appendix \ref{app:carroll sources} with $\beta= e^k \delta k_k$)
\begin{equation}
   \delta \mathcal C_i{}^j
    =
    k_i\,\delta n^j
    +
    \tfrac12\bigl(q^{kl}\delta q_{kl}\bigr)\Pi_i{}^j
    -
    \beta\,e_in^j .
    \label{eq:carrollian-source-2d}
\end{equation}
The variation of the renormalised generating functional
\(W\equiv W_{\rm AFS}[\,\mathcal C,\lambda\,]\) is
\begin{equation}
    \delta W
    =
    \int_{\mathscr I^+}\eta\,
    \left[
        \left\langle T_j{}^i\right\rangle_{\lambda}\delta \mathcal C_i{}^j
        +
        \sum_A\left\langle\mathcal O_A\right\rangle_{\lambda}\delta\lambda_A
    \right],
    \label{eq:generating-functional-variation}
\end{equation}
with \(\langle\,\cdot\,\rangle_{\lambda}\) a one-point function at finite
sources. The connected correlators are generated by repeated differentiation
with respect to \(\lambda_A\). More explicitly, one point function in the presence of the source can be written as
\begin{equation}
    \langle \mathcal O(u,\phi)\rangle_\lambda
    =
    \frac{\delta W_{\rm AFS}}{\delta \lambda(u,\phi)},
    \qquad
    \langle T_j{}^i(u,\phi)\rangle_\lambda
    =
    \frac{\delta W_{\rm AFS}}{\delta \mathcal C_i{}^j(u,\phi)}.
    \label{eq:one-point-functions}
\end{equation}
The latter is the mixed-index Brown--York stress tensor of the null boundary. This can be shown to match with the variational principle we used to find the holographic stress tensor (see \cite{Bhambure:2024ftz,Chandrasekaran:2021hxc,Parattu:2015gga}). 
The four projections of the Carrollian stress tensor are
\begin{equation}
    \begin{aligned}
        \mathcal P&\equiv n^j\left\langle T_j{}^i\right\rangle k_i=-\left\langle T_u{}^u\right\rangle
        &&\text{(supertranslation density)},
        \\
        \mathcal K&\equiv-e^j\left\langle T_j{}^i\right\rangle k_i=-\left\langle T_\phi{}^u\right\rangle
        &&\text{(superrotation density)},
        \\
        \mathcal S&\equiv e^j\left\langle T_j{}^i\right\rangle e_i=\left\langle T_\phi{}^\phi\right\rangle
        &&\text{(spatial stress)},
        \\
        \mathcal B&\equiv-n^j\left\langle T_j{}^i\right\rangle e_i=\left\langle T_u{}^\phi\right\rangle
        &&\text{(energy flux / boost response)} ,
    \end{aligned}
    \label{eq:density-definitions}
\end{equation}
which gives
\begin{equation}
    \left\langle T_i{}^i\right\rangle=\mathcal S-\mathcal P .
    \label{eq:stress-decomposition}
\end{equation}
%\vaishnavi{should it be positive $\mathcal{K}$ term}
%\hare{Is there is sign problem here?}

\subparagraph{Generating functional and Carrollian responses:}
% \begin{equation}
%     \langle \mathcal O(u,\phi)\rangle_\lambda
%     =
%     \frac{\delta W_{\rm AFS}}{\delta \lambda(u,\phi)},
%     \qquad
%     \langle T_j{}^i(u,\phi)\rangle_\lambda
%     =
%     \frac{\delta W_{\rm AFS}}{\delta \mathcal C_i{}^j(u,\phi)}.
%     \label{eq:one-point-functions}
% \end{equation}

Inserting \eqref{eq:carrollian-source-2d} into \eqref{eq:generating-functional-variation} organises the variation by geometric data as
\begin{equation}
    \delta W
    =
    \int_{\mathscr I^+}\eta\,
    \left[
        -\mathcal V_j\,\delta n^j
        +
        \tfrac12\,\mathcal T^{kl}\,\delta q_{kl}
        +
        \mathcal B^{\,k}\,\delta k_k
        +
        \sum_A\left\langle\mathcal O_A\right\rangle_{\lambda}\delta\lambda_A
    \right],
    \label{eq:three-source-variation}
\end{equation}
with
\begin{equation}
    \mathcal V_j\equiv-\left\langle T_j{}^i\right\rangle k_i
    =\mathcal P\,k_j+\mathcal K\,e_j,
    \qquad
    \mathcal T^{kl}=\mathcal S\,q^{kl},
    \qquad
    \mathcal B^{\,k}=\mathcal B\,e^k .
    \label{eq:responses}
\end{equation}
The supertranslation and superrotation densities are the response to rescaling and tilting the null generator; the response to
\(\delta q_{ij}\) is the spatial stress \(\mathcal S\); the response to
\(\delta k_i\) is the energy flux \(\mathcal B\). Consequently the
\(\mathrm{BMS}_3\) current constructed below is the contraction of the
symmetry generator with the momentum conjugate to the generator itself. For AFS without any sources or fluxes, we have found
\begin{equation}
    \mathcal P=\frac{ M(\phi)}{16\pi G_N},
    \qquad
    \mathcal K =\frac{ N(\phi)}{8\pi G_N},
    \label{eq:density-normalisation}
\end{equation}
% \vaishnavi{for K it should be 1/8piG ? And we used regular M,N not mathcal M,N for Bondi metric, is this supposed to be different?}
with \( M, N\) the aspects appearing in Bondi metric. The other components are zero.

\subparagraph{Matter flux through \(\mathscr I^+\):}

The boundary theory at \(\mathscr I^+\) is not closed. The bulk matter reaching null infinity carries flux. We denote the leak term by
\(\mathcal F^{\rm out}_j\). This is fixed by bulk Einstein's equations with matter as the matter flux decreases the Bondi mass.
\begin{equation}
    \mathcal F^{\rm out}_j
    =
    -\lim_{r\to\infty}\left(r^{\,p}\,T^{(m)}_{ji}\,n^i\right)
    \label{eq:matter-flux-definition}
\end{equation}
with \(p\) dictated by the fall-off conditions (see \cite{Cotler:2024cia} for more details about the matter fluxes).

\paragraph{Diffeomorphism Ward identity:}
The generating function \(W_{\rm AFS}\) remains invariant under diffeomorphisms generated by vector field \(\xi=\xi^i\partial_i\). The Carrollian data
\(\mathcal C_i{}^j\) transforms by the Lie derivatives of
\((q_{ij},n^i,k_i)\) and the scalar sources  also transform as
\(\delta_\xi\lambda_A=\xi^i\partial_i\lambda_A\). Hence, the invariance condition gives rise to the Ward identity
\begin{equation}
    \mathcal D_i\left\langle T_j{}^i\right\rangle_{\lambda}
    +
    \sum_A\left\langle\mathcal O_A\right\rangle_{\lambda}\partial_j\lambda_A
    =
    \left\langle\mathcal F^{\rm out}_j\right\rangle_{\lambda}.
    \label{eq:complete-diffeomorphism-ward}
\end{equation}
Here \(\mathcal D_i\) is the Carrollian connection on $\mathscr{I}^+$ \footnote{See appendix \ref{app:connection} for more details about the explicit form of this covariant derivative.} Using
\eqref{eq:density-definitions},
\begin{equation}
    \mathcal D_i\left\langle T_u{}^i\right\rangle_{\lambda}
    =-\partial_u\mathcal P,
    \qquad
    \mathcal D_i\left\langle T_{\phi}{}^i\right\rangle_{\lambda}
    =-\partial_u\mathcal K,
    \label{eq:carroll-divergence-flat}
\end{equation}
Here, we have restricted the stress tensor to the AFS case where only $\mathcal{P}$ and $\mathcal{K}$ are nonvanishing. Then the Ward identity can be written as
\begin{equation}
    \begin{aligned}
        \partial_u\mathcal P
        &=
        \sum_A\left\langle\mathcal O_A\right\rangle_{\lambda}\partial_u\lambda_A
        -\left\langle\mathcal F^{\rm out}_u\right\rangle_{\lambda},
        \\[2mm]
        \partial_u\mathcal K
        &=\sum_A\left\langle\mathcal O_A\right\rangle_{\lambda}\partial_\phi\lambda_A
        -\left\langle\mathcal F^{\rm out}_\phi\right\rangle_{\lambda}.
    \end{aligned}
    \label{eq:complete-bondi-balance}
\end{equation}
%\vaishnavi{how do u derivatives make sense when its purely phi functions} \vaishnavi{okay but 4.15 specialises to no source right? should that just be zero} \hare{here we have sources+fluxes as well, that is why. We can turn off the sources, but matter can escape to null infinity, that is why. Bondi metric is valid only in the no sources, but if we have things going outside the null infinity like matter, then masses needs to be defined with u as well, as it decrese the mass. }\vaishnavi{Okay I have some questions, lets talk today. I am at department}
Here we have not discussed the possible Weyl and Carrollian boost anomalies (see \cite{Hartong:2025jpp,Hartong:2026rbr} for more details about anomalies). At vanishing boundary sources and on the Bondi-induced frame, the first line of \eqref{eq:complete-bondi-balance} is the Bondi mass-loss law and the second is the angular-momentum loss law \cite{Fiorucci:2025twa}:
\begin{equation}
\begin{split}
    \partial_u P(u,\phi)
    & =
    -\left\langle\mathcal F^{\rm out}_u(u,\phi)\right\rangle,
     \\
    \partial_u \mathcal{K}(u,\phi) & = - \left\langle\mathcal F^{\rm out}_\phi(u,\phi)\right\rangle
    \label{eq:mass-loss-with-matter} \\
\end{split}
\end{equation}
Note that here we are considering matter flux so the supertranslation and superrotation densities $\mathcal{P},\mathcal{K}$ are not just $M(\phi),N(\phi)$ anymore. 

\paragraph{Memory and soft graviton theorem:}
Instead of arbitrary diffeomorphisms $\xi$, if we use the generators of BMS symmetry, then the identity \eqref{eq:complete-bondi-balance} can be recast into the three-dimensional soft graviton theorem of \cite{Cotler:2024cia}. The memory effect can also be written by integrating the Ward identity across null infinity. This gives the permanent shift in boundary data $\Delta\mathcal P(\phi),\Delta\mathcal K(\phi)$.

\paragraph{From transformation laws to the current algebra:}
Under a supertranslation \(T(\phi)\) and superrotation  \(R(\phi)\), the improved densities transform anomalously, %\vaishnavi{we need to explain what are the $\mathcal{P},\mathcal{J}$ we are using here and if they relate to the brown York charges we calculated earlier.} \vaishnavi{Also need a citation or eq ref for the following tranformations}
%\hare{These are components of T1 and T2- see eq 3.13, including the taylor expanded term, no need for reference, instead of $\mathcal{J}$ use $\mathcal{K}$. hence $\mathcal{K}=T1$ and $\mathcal{P}=T2$. For ASF it is written above. 4.10 }
\begin{equation}
    \delta_{T,R}\mathcal P
    =R\mathcal P'+2R'\mathcal P-\frac{c_M}{24\pi}\,R''',
    \label{eq:P-transform-universal}
\end{equation}
\begin{equation}
    \delta_{T,R}\mathcal K
    =R\mathcal K'+2R'\mathcal K+T\mathcal P'+2T'\mathcal P
    -\frac{c_L}{24\pi}\,R'''-\frac{c_M}{24\pi}\,T''',
    \label{eq:J-transform-universal}
\end{equation}
a prime denoting \(\partial_\phi\) and the third-derivative terms being the central
anomalies.  The transformations are generated canonically by the smeared charges
\begin{equation}
    P[T]=\int_0^{2\pi}\! d\phi\,T\,\mathcal P,
    \qquad
    K[R]=\int_0^{2\pi}\! d\phi\,R\,\mathcal K,
    \qquad
    \delta_{T,R}\mathcal A(\phi)=\big\{\mathcal A(\phi),\,P[T]+K[R]\big\},
    \label{eq:variation-is-bracket}
\end{equation}
so \eqref{eq:P-transform-universal}-\eqref{eq:J-transform-universal} fix every
Poisson bracket among \(\mathcal P,\mathcal K\).  The local brackets follow by
undoing the smearing with
\begin{equation}
    \int_0^{2\pi}\! d\phi'\,f(\phi')\,\partial_{\phi'}^{\,n}\delta(\phi-\phi')
    =(-1)^n f^{(n)}(\phi).
    \label{eq:delta-derivative-identity}
\end{equation}

\emph{(i) \(\{\mathcal P,\mathcal P\}=0\).}  Acting on \(\mathcal P\) with a pure
supertranslation gives, by \eqref{eq:variation-is-bracket},
\(\{\mathcal P(\phi),P[T]\}=\delta_T\mathcal P(\phi)\).  But
\eqref{eq:P-transform-universal} has no \(T\)-dependent term,
\(\delta_T\mathcal P=0\); since this holds for arbitrary \(T\), the bracket
vanishes pointwise.  Physically, the supermomenta form the abelian ideal of
\(\mathrm{BMS}_3\) and respond only to superrotations.

\emph{(ii) \(\{\mathcal K,\mathcal P\}\).}  A pure superrotation acting on
\(\mathcal P\) gives
\(\delta_R\mathcal P(\phi)=\{\mathcal P(\phi),K[R]\}
=-\int d\phi'\,R(\phi')\,\{\mathcal K(\phi'),\mathcal P(\phi)\}\).
We postulate the most general local structure and match to
\eqref{eq:P-transform-universal} via delta function identity \eqref{eq:delta-derivative-identity}. Using integration by parts, we have
\begin{equation}
    \big\{\mathcal K(\phi),\mathcal P(\phi')\big\}
    =2\mathcal P(\phi)\,\partial_\phi\delta(\phi-\phi')
    +\mathcal P'(\phi)\,\delta(\phi-\phi')
    -\frac{c_M}{24\pi}\,\partial_\phi^{3}\delta(\phi-\phi').
    \label{eq:JP-current-algebra}
\end{equation}
Equivalently, the \(T\)-sector of
\eqref{eq:J-transform-universal} is reproduced by
\(\{\mathcal K(\phi),P[T]\}\) with the same coefficient, a nontrivial consistency
check.

\emph{(iii) \(\{\mathcal K,\mathcal K\}\).}  Acting with a superrotation on
\(\mathcal K\), only the \(R\)-part of \eqref{eq:J-transform-universal} contributes,
\(\delta_R\mathcal K=R\mathcal K'+2R'\mathcal K-\tfrac{c_L}{24\pi}R'''\).
The identical matching fixes
\begin{equation}
    \big\{\mathcal K(\phi),\mathcal K(\phi')\big\}
    =2\mathcal K(\phi)\,\partial_\phi\delta(\phi-\phi')
    +\mathcal K'(\phi)\,\delta(\phi-\phi')
    -\frac{c_L}{24\pi}\,\partial_\phi^{3}\delta(\phi-\phi'),
    \label{eq:JJ-current-algebra}
\end{equation}
now with anomaly \(c_L\).  In each bracket the coefficient \(2\) of
\(\mathcal O\,\partial_\phi\delta\) is the conformal weight of the weight-two
densities, \(\mathcal O'\,\delta\) is the transport term, and
\(\partial_\phi^{3}\delta\) is the central extension whose coefficient equals the
anomaly in the corresponding transformation law. 

% Together with
% \begin{equation}
%     \big\{\mathcal P(\phi),\mathcal P(\phi')\big\}=0
%     \label{eq:PP-current-algebra}
% \end{equation}
% this is the İnönü--Wigner-contracted structure of \(\mathrm{BMS}_3\): the
% supermomenta commute, their bracket with the superrotations carries \(c_M\), and
% the superrotations close on a Virasoro subalgebra with central charge \(c_L\).

\paragraph{Mode algebra:}
Using these Poisson brackets, we can find the mode algebra as well. The Fourier modes
\begin{equation}
    M_n=\int_0^{2\pi}\! d\phi\,e^{in\phi}\,\mathcal P(\phi),
    \qquad
    L_n=\int_0^{2\pi}\! d\phi\,e^{in\phi}\,\mathcal K(\phi),
\end{equation}
and \([\,\cdot\,,\cdot\,]=i\{\,\cdot\,,\cdot\,\}\), the brackets gives the centrally
extended \(\mathrm{BMS}_3\) algebra
\begin{align}
    [L_m,L_n]&=(m-n)L_{m+n}+\frac{c_L}{12 }\,m(m^2-1)\,\delta_{m+n,0},
    \\[2pt]
    [L_m,M_n]&=(m-n)M_{m+n}+\frac{c_M}{12 }\,m(m^2-1)\,\delta_{m+n,0},
    \\[2pt]
    [M_m,M_n]&=0,
\end{align}
\section{The BMS$_3$ boundary graviton action}
\label{sec:null-boundary-bms-schwarzian}

We now show that the null-boundary term of the action in 3-dimensional flat gravity gives rise to the boundary gravitons' action. The result coincides with the boundary action obtained by first-order/Chern-Simons reduction in \cite{Barnich:2017jgw,Merbis:2019wgk}
and by Cotler et al.\ \cite{Cotler:2024cia}.

\paragraph{Boundary term:}
For a null hypersurface $\mathcal N$ with generator $n^a$, transverse density $\sqrt q$, inaffinity $\kappa$, and expansion $\Theta$, we know 
\begin{equation}
   n^b\nabla_bn^a=\kappa\,n^a,
    \qquad
    \Theta=\frac{1}{\sqrt q}\,\mathcal L_n\sqrt q .
\end{equation}
The null-boundary contribution of the bulk action \cite{Parattu:2015gga,Chandrasekaran:2021hxc} is
\begin{equation}
    I_{\mathcal N}
    =
    \frac{1}{8\pi G_N}
    \int_{\mathcal N} du\,d\phi\,
    \sqrt q\,(\kappa+\Theta).
\end{equation}

The 3d asymptotically flat spacetime in Bondi gauge is given by \eqref{eq:Bondi metric1} which we rewrite here:
\begin{equation}
    ds^2
    =
    M(\phi)\,du^2
    -2\,du\,dr
    +2\left[
        N(\phi)+\frac{u}{2}\partial_\phi M(\phi)
    \right]du\,d\phi
    +r^2d\phi^2 .
\end{equation}
The normal which gives a bulk foliation with the limit of being the null normal at $\mathcal{I}^+$ is
\begin{equation}
    n_\mu dx^\mu
    =
    dr-\frac{M(\phi)}{2}\,du
    -\frac{u}{2}\partial_\phi M(\phi)\,d\phi ,\quad n^\mu
    =
    \left\{
        -\frac{M}{2}
        +\frac{N\left(u\,\partial_\phi M+2N\right)}{2r^2},
        -1,
        \frac{N}{r^2}
    \right\}.
\end{equation}
% whose raised form is
% \begin{equation}
%     n^\mu
%     =
%     \left\{
%         -\frac{M}{2}
%         +\frac{N\left(u\,\partial_\phi M+2N\right)}{2r^2},
%         -1,
%         \frac{N}{r^2}
%     \right\}.
% \end{equation}
%Its norm is $  n^\mu n_\mu=\frac{N^2}{r^2}$ which becomes null at $\mathscr I^+$.

The asymptotic null generator is therefore
\begin{equation}
    \lim_{r\to\infty}n^\mu
    =
    \left\{
        -\frac{M}{2},-1,0
    \right\},
\end{equation}
and is affinely parametrized,
\begin{equation}
   n^\nu\nabla_\nu n^\mu=0 \rightarrow \kappa=0.
\end{equation}

On a cut of $\mathscr I^+$ we get $\sqrt q=r$ which gives
\begin{equation}
    \Theta
    =
    \frac{1}{r}n^\mu\partial_\mu r
    =
    -\frac{M(\phi)}{2r}.
\end{equation}
% Keeping the full asymptotic normal gives only a subleading correction,
% \begin{equation}
%     \frac{1}{r}n^\mu\partial_\mu r
%     =
%     -\frac{M}{2r}
%     +
%     \frac{N\left(u\,\partial_\phi M+2N\right)}{2r^3}.
% \end{equation}
Thus
\begin{equation}
    \sqrt q\,\Theta
    =
    -\frac{M(\phi)}{2}
    +O\left(\frac{1}{r^2}\right),
\end{equation}
and the null-boundary term becomes
\begin{align}
    I_{\Theta}^{\rm bdry}
    &=
    \lim_{r\to\infty}
    \frac{1}{8\pi G_N}
    \int_{\mathscr I^+}du\,d\phi\,
    \sqrt q\,\Theta= -\frac{1}{16\pi G_N}
    \int_{\mathscr I^+}du\,d\phi\,M(\phi)
    \nonumber\\
    &=
    -\int_{\mathscr I^+}du\,d\phi\,
    \mathcal P(\phi),
\end{align}
where
\begin{equation}
    \mathcal P(\phi)=\frac{M(\phi)}{16\pi G_N}.
\end{equation}
This gives
\begin{equation}
    I_{\Theta}^{\rm bdry}
    =
    -\Delta u\,H,
    \qquad
    H=\int_0^{2\pi} d\phi\,\mathcal P(\phi),
\end{equation}
Thus the null-boundary contribution is the Hamiltonian term
associated with the zero mode of the supermomentum. We elaborate on this point below.

\paragraph{BMS$_3$ dressing and the Schwarzian:}
Consider a finite \(\mathrm{BMS}_3\) transformation
\begin{equation}
    \phi\to f(\phi),\qquad
    u\to f'(\phi)\,u+g(\phi),
    \label{eq:finite-bms}
\end{equation}
with \(f\in \mathrm{Diff}^+(S^1)\).  The superrotation part acts on the mass
aspect by the coadjoint transformation \cite{Barnich:2015uva}
\begin{equation}
    M_f(\phi)
    =
    f'(\phi)^2 M(f(\phi))-2\{f,\phi\},
    \qquad
    \{f,\phi\}
    =
    \frac{f'''}{f'}
    -\frac32\left(\frac{f''}{f'}\right)^2 .
    \label{eq:mass-coadjoint}
\end{equation}

For the Minkowski spacetime, we have $M=-1$, hence using the Schwarzian chain-rule we identify
\begin{equation}
    \left\{\tan\frac{f}{2},\phi\right\}
    =
    \{f,\phi\}+\frac12 f'^2,
    \label{eq:tan-schwarzian-identity}
\end{equation}
and we obtain
\begin{equation}
    I_{\Theta}^{\rm mink}
    =
    \frac{1}{8\pi G_N}\int du d\phi
    \left\{\tan\frac{f(\phi)}{2},\phi\right\}
     .
    \label{eq:supermomentum-schwarzian}
\end{equation}
Here $\mathcal{P}=- \frac{1}{8 \pi G}\left\{\tan\frac{f(\phi)}{2},\phi\right\}$. The Schwarzian is invariant under $PSL(2,\mathbb R)$ M\"obius transformations of $\tan(f/2)$. This is the stabiliser of $\mathcal P_{\rm vac}=-1/16\pi G_N$. Hence, the superrotation mode lives in $\mathrm{Diff}(S^1)/PSL(2,\mathbb R)$. This is only half the story: We now denote the supertranslation coordinate on the $\mathrm{BMS}_3$ orbit by $\alpha$ (related to the finite supertranslation $g$ above by a convention-dependent\cite{Cotler:2024cia}, $f$-dependent redefinition). Thus, a point on the vacuum orbit is parametrized by the pair $(f,\alpha)$, where $f$ is the superrotation coordinate and $\alpha$ is the supertranslation coordinate. Hence, the boundary-graviton phase space is the vacuum coadjoint orbit
\begin{equation}
    \mathrm{BMS}_3/ISO(2,1),\qquad ISO(2,1)=PSL(2,\mathbb R)\ltimes\mathbb R^3 ,
\end{equation}
which, supertranslations forming an abelian ideal, fibers over the superrotation quotient as $T^{*}\big(\mathrm{Diff}(S^1)/PSL(2,\mathbb R)\big)$.

Consequently, \eqref{eq:supermomentum-schwarzian} is not the full boundary action but only the Hamiltonian. Promoting $f\to f(u,\phi)$ and adding the presymplectic potential of the
orbit gives the first-order action
\begin{equation}
    I_{\rm bdy}[\alpha,f]
    =-\int du\,d\phi\,\Big(\alpha\,\partial_u\mathcal P[f]+\mathcal P[f]\Big),
    \qquad
    \omega=\int d\phi\,\delta\alpha\wedge\delta\mathcal P ,
\end{equation}
in agreement with \cite{Merbis:2019wgk,Cotler:2024cia}. Varying $\alpha$ imposes $\partial_u\mathcal P=0$, while varying with respect to $f$ gives equation for $\mathcal{J}[\alpha,f]$ as $\partial_u\mathcal J=\partial_\phi\mathcal P$
with $\mathcal J=\alpha\partial_\phi\mathcal P+2\mathcal P\partial_\phi\alpha
-\tfrac{1}{8\pi G_N}\partial_\phi^3\alpha$. These are the flat-space Bondi equations. And it is this term that supplies the phase-space structure reproducing the centrally extended BMS algebra \cite{Barnich:2015uva}.

\appendix

\section*{Acknowledgements}
We would like to thank E. Caceres and A. Karch for enlightening discussions. The work of H.K.  is supported in part by CNS Spark Grant 2025.
\appendix

\section{A review of free Carrollian scalars}
\label{app:free-scalar}

\subsection*{Carrollian CFTs}
\label{sec:carrCFT}
 
% A \emph{Carrollian conformal field theory} in two dimensions (CCFT$_2$), equivalently a
% BMS$_3$-invariant field theory, is a quantum field theory whose global spacetime symmetry is
% the conformal Carrollian / BMS$_3$ algebra \eqref{eq:BMSalgebra}. There theories are not Lorentz invariant as these are defined on Carrollian background. BMS symmetries acts on spatial direction $\sigma$ and a Carrollian
% time $\tau$ as
% \begin{equation}
%     \sigma \to f(\sigma), \qquad \tau \to f'(\sigma)\,\tau + g(\sigma),
%     \label{eq:CCFTmap}
% \end{equation}
% where $f$ is a reparametrization of the circle (superrotation) and $g$ a $\sigma$-dependent
% shift of the Carrollian time (supertranslation). 

Carrollian CFTs can arise in two ways. Intrinsically, a CCFT$_2$ may be defined (similar to the conformal bootstrap) purely from the representation theory of BMS or Carroll group without reference to any Lagrangian \cite{Bagchi:2009pe,Hao:2021urq}. Alternatively, they appear as the $c\to 0$ (ultra-relativistic) limit of relativistic CFT$_2$. This limit is not unique. It depends on how the fields are scaled before the contraction.  One obtains inequivalent \emph{electric} or \emph{magnetic} Carrollian theories \cite{Bagchi:2022eav, Ruzziconi:2026bix,Henneaux:2021yzg}. For a scalar, the electric theory retains only the time-derivative term, $\mathcal{L}=\tfrac12(\partial_u\Phi)^2$, which makes the ultralocal character manifest. See \cite{Marotta:2025qjh,Hao:2025naz,deBoer:2021jej} for recent work in Carrollian CFTs and Carrollian limits.

Local operators are organized, as in CFT$_2$, into primaries and their descendants. There are \emph{two} quantum numbers: the conformal weight $\Delta$ and the boost charge
$\chi$, defined as the eigenvalues under the zero modes of the generator
\begin{equation}
    [L_0,\mathcal{O}] = \Delta\,\mathcal{O}, \qquad
    [M_0,\mathcal{O}] = \chi\,\mathcal{O}, \qquad
    [L_n,\mathcal{O}] = [M_n,\mathcal{O}] = 0 \ \ (n>0).
    \label{eq:CCprimary}
\end{equation}
and the descendants generated by $L_{-n},M_{-n}$.

\subsection{The three actions and the field transformations}
\label{app:actions}
As time and space behave differently in this paradigm, we separate the components of the metric.
On the flat Carrollian cylinder with the zweibeins $(\tau^\mu,e^\mu_1)$, we contract $\partial_\mu\Phi$ with zweibeins. It yields three inequivalent two-derivative actions for
a single massless scalar $\Phi(u,\phi)$ 
\cite{Bagchi:2022eav} as
\begin{equation}
\begin{split}
    & S_t=\int du\,d\phi\,(\partial_u\Phi)^2, \\
    & S_m=\int du \, d\phi\, \Big[\partial_\phi\Phi\,\partial_u\Phi
        +e^{u}_1(\partial_u\Phi)^2\Big],\\
    & S_{sp}=\int du\,d\phi\,\big(\partial_\phi\Phi+e^{u}_1\partial_u\Phi\big)^2.
    \label{eq:three-actions}
\end{split}
\end{equation}
These are called the timelike, mixed-derivative, and spacelike actions (or electric, mixed and magnetic actions) respectively. The boost zweibein $e^{u}_1$ is left unfixed by the flat-Carroll gauge choice.

The BMS$_3$ transformation is generated by the conformal
Carroll Killing vector
\begin{equation}
    \xi^{\phi}=\varepsilon\,\mathcal{R}(\phi),\qquad
    \xi^{u}=\varepsilon (u\,\mathcal{R}'(\phi)+\mathcal{T}(\phi)),
    \label{eq:killing}
\end{equation}
 where primes denote $\partial_\phi$ throughout. The derivatives are:
\begin{equation}
    \partial_u\xi^{u}=\varepsilon \mathcal{R}',\qquad
    \partial_\phi\xi^{\phi}=\varepsilon \mathcal{R}',\qquad \partial_\phi\xi^{u}=\varepsilon (u\mathcal{R}''+\mathcal{T}')\equiv\Xi,\qquad
    \partial_u\xi^{\phi}=0.
    \label{eq:killing-derivs}
\end{equation}
We write $\mathcal{D}_{\xi} X \equiv \xi^\rho \partial_\rho X= \xi^u \partial_u X + \xi^\phi \partial_\phi X$
for the transport (Lie) piece. Since $\Phi$ is a scalar field with BMS-Weyl weight zero,
$\delta\Phi=-\xi^\rho\partial_\rho\Phi$, and because $\delta$ commutes with
$\partial_\mu$,
\begin{align}
    \delta(\partial_u\Phi)
        &=-\varepsilon \mathcal{R}' \partial_u\Phi-\mathcal{D}_{\xi}(\partial_u\Phi),
        \label{eq:dPhi-u}\\
    \delta(\partial_\phi\Phi)
        &=-\Xi\partial_u\Phi-\varepsilon \mathcal{R}'\partial_\phi\Phi
          -\mathcal{D}_{\xi}(\partial_\phi\Phi).
        \label{eq:dPhi-phi}
\end{align}
while the unfixed zweibein transforms as \cite{Bagchi:2022eav}:,
\begin{equation}
    \delta e^{u}_1=-\mathcal{D}_{\xi} e^{u}_1+\partial_\phi\xi^{u}
                 =-\mathcal{D}_{\xi} e^{u}_1+\Xi.
    \label{eq:de1}
\end{equation}
The above three actions are invariant off-shell provided $e_1^u$ transforms.
\subsection{Stress tensors}
\label{app:stress}

The stress tensors corresponding to the three actions \eqref{eq:three-actions} are defined as
\begin{equation}
    T^\mu{}_\nu=\frac{e^\mu_A}{2e}\frac{\delta S}{\delta e^\nu_A}
    \label{eq: strees tensor for scalar}
\end{equation}
and they are traceless ($T^{u}{}_{u}=-T^{\phi}{}_{\phi}$) as expected from the Carrollian conformal symmetry.\\

\paragraph{An Aside:}
The above formula for the stress tensor and the Brown-York stress tensor discussed in this article can be compared as follows. 
The stress tensor of an intrinsic Carrollian scalar theory may be defined in direct analogy with the stress tensor of a relativistic field
theory.  We first couple the scalar field to a general Carrollian
background as
\begin{equation}
    S_\Phi=S_\Phi[\Phi;q_{\mu\nu},n^\mu],
    \qquad
    q_{\mu\nu}n^\nu=0 .
\end{equation}
For variations preserving the Carrollian constraint, the variation takes
the form
\begin{equation}
    \delta S_\Phi
    =
    \int_{\mathcal N}\eta\,
    \left(
        E_\Phi\,\delta\Phi
        +p_\Phi^{\mu\nu}\,\delta q_{\mu\nu}
        +p^\Phi_\mu\,\delta n^\mu
    \right)
    +\text{boundary terms},
    \label{eq:carrollian-scalar-variation}
\end{equation}
where
\begin{equation}
    p_\Phi^{\mu\nu}
    \equiv
    \frac{1}{\eta}\frac{\delta S_\Phi}{\delta q_{\mu\nu}},
    \qquad
    p^\Phi_\mu
    \equiv
    \frac{1}{\eta}\frac{\delta S_\Phi}{\delta n^\mu}.
\end{equation}
Under a boundary diffeomorphism,
\(\delta_\xi q_{\mu\nu}=\mathcal L_\xi q_{\mu\nu}\) and
\(\delta_\xi n^\mu=\mathcal L_\xi n^\mu\), isolating the terms
proportional to derivatives of \(\xi^\mu\) gives the mixed-index
Carrollian stress tensor
\begin{equation}
    T^\mu{}_\nu
    =
    2p_\Phi^{\mu\rho}q_{\rho\nu}
    -n^\mu p^\Phi_\nu .
    \label{eq:carrollian-scalar-stress}
\end{equation}
The factor of two follows from the variation of the symmetric tensor
\(q_{\mu\nu}\). This is what we get from null boundary construction involving inaffinity and expansion \cite{Parattu:2015gga,Chandrasekaran:2021hxc}.

In the zweibein formulation, the Carrollian fields are
\begin{equation}
    n^\mu=e^\mu{}_0,
    \qquad
    q_{\mu\nu}=e^1{}_\mu e^1{}_\nu,
    \qquad
    e\equiv\det(e^A{}_\mu).
\end{equation}
Using
\begin{equation}
    \delta e^A{}_\mu
    =
    -e^A{}_\rho e^B{}_\mu\,\delta e^\rho{}_B ,
\end{equation}
the chain rule gives
\begin{equation}
    \frac{e^\mu{}_A}{e}
    \frac{\delta S_\Phi}{\delta e^\nu{}_A}
    =
    -2p_\Phi^{\mu\rho}q_{\rho\nu}
    +n^\mu p^\Phi_\nu
    =
    -T^\mu{}_\nu .
\end{equation}
Hence for the inverse-zweibein convention
\begin{equation}
    T^\mu{}_{\nu\,(\Phi)}
    =
    \frac{e^\mu{}_A}{2e}
    \frac{\delta S_\Phi}{\delta e^\nu{}_A},
\end{equation}
one obtains
\begin{equation}
    T^\mu{}_\nu
    =
    -2T^\mu{}_{\nu\,(\Phi)} .
\end{equation}
This identification assumes that the action depends on the zweibeins
only through \(q_{\mu\nu}\) and \(n^\mu\).  If additional independent
Carroll-frame data are present; their conjugate response must be
included as well.  The flat Carrollian background may be imposed
only after the variational derivatives have been evaluated.\\

For Carroll scalar theories, the explicit components are \cite{Bagchi:2022eav}:
\begin{align}
    \text{timelike:}\quad
    & T_2 \equiv T^u{}_{u}=\tfrac12(\partial_u\Phi)^2,\quad
     T_1\equiv T^{u}{}_{\phi}=\partial_u\Phi \partial_\phi\Phi,\quad
     T^{\phi}{}_{u}=0,
     \label{eq:stress-timelike}\\
    \text{mixed:}\quad
    & T_2 \equiv \tfrac12 e^{u}_1(\partial_u\Phi)^2,\quad T_1 \equiv \tfrac12(\partial_\phi\Phi)^2+e^{u}_1\partial_\phi\Phi\partial_u\Phi,\quad
     T^{\phi}{}_{u}=\tfrac12(\partial_u\Phi)^2,
     \label{eq:stress-mixed}\\
    \text{spacelike:}\quad
    & T_2 \equiv \tfrac12 \left[ -(\partial_\phi\Phi)^2+(e^{u}_1)^2(\partial_u\Phi)^2 \right],\quad T_1 \equiv e^{u}_1\partial_\phi\Phi ( \partial_\phi\Phi + e^{u}_1 \partial_u\Phi ),\nonumber \\ 
    & T^{\phi}{}_{u} = \partial_u\Phi (\partial_\phi\Phi+e^{u}_1 \partial_u\Phi),
     \label{eq:stress-spacelike}
\end{align}
with the identifications we have used throughout the article \eqref{eq:T1-T2-identification}. Note that only the timelike theory exhibits the Carroll-boost invariance ($T^{\phi}{}_{u}=0$) off-shell. The other two have $T^{\phi}{}_{u}\neq0$ but it becomes zero assuming the equations of motion for $e_1^u$ \cite{Bagchi:2022eav}. If the unfixed frame component $e_1^u$ is treated dynamically, its equation of motion sets $T^{\phi}{}_{u}=0$. In a fixed-background formulation, this condition does not follow from the scalar equation of motion; an appropriate improvement or additional Ward identity is required.

\subsection*{Transformation of the stress tensors}
\label{app:transform}

\paragraph{Timelike:}
Using \eqref{eq:dPhi-u} in $T_2$:
\begin{equation}
    \delta T_2=\partial_u\Phi\,\delta(\partial_u\Phi)
    =-\varepsilon \mathcal{R}'(\partial_u\Phi)^2-\partial_u\Phi \mathcal{D}_\xi (\partial_u\Phi)
    =-\mathcal{D}_\xi T_2 - 2 \varepsilon \mathcal{R}' T_2,
    \label{eq:dT2-timelike}
\end{equation}
where we used $\partial_u\Phi\,\mathcal{D}_\xi(\partial_u\Phi)=\mathcal{D}_\xi\big(\tfrac12(\partial_u\Phi)^2\big)=\mathcal{D}_\xi T_2$
and $(\partial_u\Phi)^2=2T_2$. Using \eqref{eq:dPhi-u} and \eqref{eq:dPhi-phi} in $T_1$, the transport terms combine into $-\mathcal{D}_\xi T_1$, the two $\varepsilon \mathcal{R}'$ terms give
$-2\varepsilon \mathcal{R}' T_1$, and the $-\Xi(\partial_u\Phi)^2=-2\Xi T_2$ term remains giving:
\begin{equation}
    \delta T_1=-\mathcal{D}_\xi T_1-2\varepsilon \mathcal{R}'\,T_1-2\Xi\,T_2.
    \label{eq:dT1-timelike}
\end{equation}

\paragraph{Mixed-derivative:}
Here $e^{u}_1$ also varies via \eqref{eq:de1}. We get
\begin{equation}
    \delta T_2=\tfrac12(\delta e^{u}_1)(\partial_u\Phi)^2
              +e^{u}_1\partial_u\Phi\,\delta(\partial_u\Phi)
    =\tfrac12(-\mathcal{D}_\xi e^{u}_1+\Xi)(\partial_u\Phi)^2
     +e^{u}_1\partial_u\Phi\big[-\varepsilon \mathcal{R}'\partial_u\Phi-\mathcal{D}_\xi(\partial_u\Phi)\big].
\end{equation}
The terms with $\mathcal{D}_\xi$ assemble into
$-\mathcal{D}_\xi T_2=-\tfrac12(\mathcal{D}_\xi e^{u}_1)(\partial_u\Phi)^2-e^{u}_1\partial_u\Phi\,\mathcal{D}_\xi(\partial_u\Phi)$,
the $\varepsilon \mathcal{R}'$ term gives $-\varepsilon \mathcal{R}'e^{u}_1(\partial_u\Phi)^2=-2\varepsilon \mathcal{R}'T_2$
and the leftover $\tfrac12\Xi(\partial_u\Phi)^2=\Xi\,T^{\phi}{}_{u}$. Hence
\begin{equation}
    \delta T_2=-\mathcal{D}_\xi T_2-2\varepsilon \mathcal{R}'\,T_2+\Xi\,T^{\phi}{}_{u}.
    \label{eq:dT2-mixed}
\end{equation}
For $T_1$ we expand $\delta T_1=\partial_\phi\Phi\,\delta(\partial_\phi\Phi) +(\delta e^{u}_1)\partial_\phi\Phi\,\partial_u\Phi +e^{u}_1(\delta\partial_\phi\Phi)\partial_u\Phi +e^{u}_1\partial_\phi\Phi(\delta\partial_u\Phi)$. Using \eqref{eq:dPhi-u}--\eqref{eq:de1} the two $\Xi\,\partial_\phi\Phi\,\partial_u\Phi$ pieces (one from $\delta\partial_\phi\Phi$ and one from $\delta e^{u}_1$) cancel and the transport pieces sum to $-\mathcal{D}_\xi T_1$. The $\varepsilon \mathcal{R}'$ pieces sum to $-2\varepsilon \mathcal{R}'T_1$ and the surviving piece is
$-\Xi e^{u}_1(\partial_u\Phi)^2=-2\Xi T_2$, giving
\begin{equation}
    \delta T_1=-\mathcal{D}_\xi T_1-2\varepsilon \mathcal{R}'\,T_1-2\Xi\,T_2,
    \label{eq:dT1-mixed}
\end{equation}
the same form as the timelike case \eqref{eq:dT1-timelike}.

\paragraph{Spacelike:}
The algebra is longer because of the $(e^{u}_1)^2$ terms, but the outcome is
identical in structure: substituting \eqref{eq:dPhi-u}--\eqref{eq:de1} into
\eqref{eq:stress-spacelike} gives
\begin{equation}
    \delta T_2=-\mathcal{D}_\xi T_2-2\varepsilon \mathcal{R}'\,T_2+\Xi\,T^{\phi}{}_{u},
    \qquad
    \delta T_1=-\mathcal{D}_\xi T_1-2\varepsilon \mathcal{R}'\,T_1-2\Xi\,T_2,
    \label{eq:dT-spacelike}
\end{equation}
now with $T^{\phi}{}_{u}=\partial_u\Phi(\partial_\phi\Phi+e^{u}_1\partial_u\Phi)$.

\subsection*{central charges}
\label{app:central charges for scalar}

As seen in \eqref{eq:dT2-timelike}--\eqref{eq:dT-spacelike}, all three stress tensors
obey the same transformation for both the components of the stress tensor. These transformations are proportional to the stress tensor components itself and there are no anomalous/non-homogeneous terms. This gives
\begin{equation}
c_L^{\rm cl}=0,\qquad c_M^{\rm cl}=0
    \label{eq:free-central-zero}
\end{equation}
for the timelike, mixed-derivative and spacelike theories alike. This conclusion does not determine the quantum central charges, which
may depend on the choice of vacuum, normal ordering, improvements and
boundary terms. The quantum version of Carrollian CFT$_2$ is explored in \cite{Hao:2021urq}, where they study the timelike action of \cite{Bagchi:2022eav}, and they find nonzero central charges as expected (specifically, they find $c_L=2,\,c_M=0$). It would be interesting to explore other types of actions and see if they give the charges we get in gravity. We hope to come back to this question in future work.\\

\section{The Carrollian contraction and the flat-space limit}
\label{app:carroll limit}
% We start with the metric defined on the plane as
% \begin{eqnarray}
%     ds^2=-\epsilon^2 dy^2+dx^2, \quad \epsilon \rightarrow 0
% \end{eqnarray}
% Then, we do the following transformation to go from the plane to the cylinder as
% \begin{eqnarray}
%     x=e^{i\sigma}, y=i\,\tau\,e^{i\sigma},
%     \label{eq: cylinder to plane map}
% \end{eqnarray}
% Then the metric on the cylinder becomes
% \begin{eqnarray}
%     ds^2= e^{2 i \sigma} (-d\sigma^2+\epsilon^2(d\tau^2+2 i \tau d\tau d\sigma-\tau^2 d\sigma^2))
% \end{eqnarray}
% As one takes the Carrollian limit $\epsilon \rightarrow 0$, we have a degenerate metric. The charges are then realized as a contour
% integrals of the stress-tensor components introduced below, 
% \begin{equation}
%     L_n = \frac{1}{2\pi i}\oint dx\,\Big(x^{n+1}\,T_1
%           + (n+1)\,x^{n}\,y\,T_2\Big),
%     \qquad
%     M_n = \frac{1}{2\pi i}\oint dx\;x^{n+1}\,T_2,
%     \label{eq:CCcharges}
% \end{equation}
% which is the Carrollian analogue of the Virasoro charge $L_n=\tfrac{1}{2\pi i}\oint x^{n+1}T(x)dx$
% \cite{Hao:2021urq}.

It is useful to recall how the BMS$_3$ generators arise from the
ultra-relativistic contraction of two Virasoro algebras.  We begin with a
relativistic two-dimensional CFT written in complex coordinates
\begin{equation}
    ds^2 = dz\,d\bar z .
\end{equation}
For the flat-space limit of AdS$_3$ gravity, we introduce cylinder coordinates
\((u,\phi)\) through
\begin{equation}
    z = e^{i(\phi+\epsilon u)},
    \qquad
    \bar z = e^{i(\epsilon u-\phi)}.
    \qquad
     .
\end{equation}
The flat-space, or Carrollian limit is
\begin{equation}
    \ell_{\rm AdS}\rightarrow \infty \quad \, \text{or}
    \qquad
    \epsilon\rightarrow 0 , \quad  \epsilon \equiv \frac{1}{\ell_{\rm AdS}}
\end{equation}
In this limit the \(u\)-direction becomes the Carrollian, or null direction. The two copies of the Virasoro generators on the plane are
\begin{equation}
    \ell_n = - z^{n+1}\partial_z,
    \qquad
    \bar\ell_n = - \bar z^{\,n+1}\partial_{\bar z}.
\end{equation}
Using
\begin{equation}
    \partial_\phi
    =
    i z\partial_z
    -
    i\bar z\partial_{\bar z},
    \qquad
    \partial_u
    =
    i\epsilon z\partial_z
    +
    i\epsilon \bar z\partial_{\bar z},
\end{equation}
one obtains
\begin{equation}
    \ell_n
    =
    \frac{i}{2}z^n
    \left(
        \partial_\phi
        +
        \frac{1}{\epsilon}\partial_u
    \right), \quad    \bar\ell_n
    =
    \frac{i}{2}\bar z^{\,n}
    \left(
        -\partial_\phi
        +
        \frac{1}{\epsilon}\partial_u
    \right).
\end{equation}

The BMS$_3$ generators are defined by the contraction
\begin{equation}
    L_n
    =
    \lim_{\epsilon\to0}
    \left(
        \ell_n-\bar\ell_{-n}
    \right),
    \qquad
    M_n
    =
    \lim_{\epsilon\to0}
    \epsilon
    \left(
        \ell_n+\bar\ell_{-n}
    \right).
\end{equation}
In the Carrollian limit, we have
\begin{equation}
    z^n
    =
    e^{in\phi}
    \left(
        1+in\epsilon u+\mathcal O(\epsilon^2)
    \right),
    \qquad
    \bar z^{-n}
    =
    e^{in\phi}
    \left(
        1-in\epsilon u+\mathcal O(\epsilon^2)
    \right).
\end{equation}
Substituting these expansions gives
\begin{equation}
    L_n
    =
    i e^{in\phi}
    \left(
        \partial_\phi
        +
        in u\,\partial_u
    \right), \quad    M_n
    =
    i e^{in\phi}\partial_u .
\end{equation}
These are the standard BMS$_3$ vector field generators on the null cylinder. These form a centrally extended BMS$_3$ algebra as
\begin{equation}
    [L_n,L_m]
    =
    (n-m)L_{n+m}
    +
    \frac{c_L}{12}n(n^2-1)\delta_{n+m,0},
\end{equation}
\begin{equation}
    [L_n,M_m]
    =
    (n-m)M_{n+m}
    +
    \frac{c_M}{12}n(n^2-1)\delta_{n+m,0},
\end{equation}
and
\begin{equation}
    [M_n,M_m]=0 .
\end{equation}

The central terms follow from the same contraction \cite{Barnich:2006av}. Suppose the parent CFT
has two Virasoro algebras,
\begin{equation}
    [\ell_n,\ell_m]
    =
    (n-m)\ell_{n+m}
    +
    \frac{c}{12}n(n^2-1)\delta_{n+m,0},
\end{equation}
and
\begin{equation}
    [\bar\ell_n,\bar\ell_m]
    =
    (n-m)\bar\ell_{n+m}
    +
    \frac{\bar c}{12}n(n^2-1)\delta_{n+m,0}.
\end{equation}
With
\begin{equation}
    L_n=\ell_n-\bar\ell_{-n},
    \qquad
    M_n=\epsilon\left(\ell_n+\bar\ell_{-n}\right),
\end{equation}
one obtains
\begin{equation}
    c_L=c-\bar c,
    \qquad
    c_M=\epsilon(c+\bar c).
\end{equation}

For Einstein gravity in AdS$_3$, the Brown--Henneaux central charges are \cite{Brown:1986nw}
\begin{equation}
    c=\bar c=\frac{3\ell_{\rm AdS}}{2G_N}.
\end{equation}
Therefore, with \(\epsilon=1/\ell_{\rm AdS}\) \cite{Barnich:2006av},
\begin{equation}
    c_L=0,
    \qquad
    c_M=\frac{3}{G_N}.
\end{equation}
% If instead one normalizes the supertranslation generator as
% \begin{equation}
%     M_n
%     =
%     \frac{1}{2\ell_{\rm AdS}}
%     \left(
%         \ell_n+\bar\ell_{-n}
%     \right),
% \end{equation}
% then the mixed central charge is correspondingly
% \begin{equation}
%     c_M=\frac{3}{2G}.
% \end{equation}
% Thus the numerical value of \(c_M\) depends on the normalization of
% \(M_n\), while the algebraic contraction itself is fixed by the flat-space
% limit.\\
Equivalently, one may map the cylinder to the Carrollian plane by
\begin{equation}
    X=e^{i\phi},
    \qquad
    Y=-iu e^{i\phi}.
\end{equation}
Then
\begin{equation}
    \partial_\phi
    =
    iX\partial_X+iY\partial_Y,
    \qquad
    \partial_u
    =
    -iX\partial_Y .
\end{equation}
Therefore the BMS$_3$ generators become
\begin{equation}
    L_n
    =
    -X^{n+1}\partial_X
    -(n+1)X^nY\,\partial_Y,
\end{equation}
and
\begin{equation}
    M_n
    =
    X^{n+1}\partial_Y .
\end{equation}
This is the plane representation of the same BMS$_3$ algebra.  The choice
\(Y=-iu e^{i\phi}\) fixes the sign convention for \(M_n\); choosing
\(Y=iu e^{i\phi}\), or equivalently \(u\to -u\), reverses the sign of
\(M_n\) without changing the algebra.

\section{Explicit evaluation of the Carrollian divergence}
\label{app:connection}

We now explicitly evaluate the projected covariant divergence of the
mixed-index stress tensor using the normal adapted to the hypersurfaces
\begin{equation}
    r-\frac{u}{2}M(\phi)=\text{constant},
\end{equation}
and derive the corresponding boundary conservation equations.

We introduce
\begin{equation}
    A(u,\phi)
    \equiv
    N(\phi)+\frac{u}{2}\partial_\phi M(\phi),
\end{equation}
in terms of which the Bondi metric reads
\begin{equation}
    ds^2
    =
    M(\phi)\,du^2
    -2\,du\,dr
    +2A(u,\phi)\,du\,d\phi
    +r^2d\phi^2 .
    \label{eq:bondi-metric-A}
\end{equation}
In the coordinate ordering \(x^a=(r,u,\phi)\), the metric and its inverse
are
\begin{equation}
    g_{ab}
    =
    \begin{pmatrix}
        0  & -1 & 0 \\
        -1 & M  & A \\
        0  & A  & r^2
    \end{pmatrix},
    \qquad
    g^{ab}
    =
    \begin{pmatrix}
        -M+\dfrac{A^2}{r^2}
        &
        -1
        &
        \dfrac{A}{r^2}
        \\[6pt]
        -1 & 0 & 0
        \\[4pt]
        \dfrac{A}{r^2}
        &
        0
        &
        \dfrac{1}{r^2}
    \end{pmatrix}.
    \label{eq:bondi-inverse-metric}
\end{equation}

The normal covector and auxiliary rigging vector are chosen as
\begin{equation}
    n_a dx^a
    =
    dr-\frac{M(\phi)}{2}\,du
    -\frac{u}{2}\partial_\phi M(\phi)\,d\phi,
    \qquad
    k^a\partial_a=-\partial_r,
    \qquad
    n_a k^a=-1.
    \label{eq:normal-rigging-divergence}
\end{equation}
Raising the index gives
\begin{equation}
    n^a\partial_a
    =
    \left[
        -\frac{M}{2}
        +
        \frac{N\left(u\,\partial_\phi M+2N\right)}{2r^2}
    \right]\partial_r
    -\partial_u
    +\frac{N}{r^2}\partial_\phi .
    \label{eq:raised-normal-divergence}
\end{equation}
Its norm is
\begin{equation}
    n^a n_a=\frac{N^2}{r^2},
\end{equation}
which becomes null at \(\mathscr{I}^+\), while remaining spacelike
at order \(1/r^{2}\) in the bulk. In particular,
\begin{equation}
    n^a\partial_a
    =
    -\frac{M}{2}\partial_r
    -\partial_u
    +O\left(\frac{1}{r^2}\right)
\end{equation}
at null infinity.

The mixed projector from the bulk tangent space to the constant
\(r-uM/2\) hypersurfaces is
\begin{equation}
    \Pi_i{}^a
    =
    \delta_i{}^a+n_i k^a .
    \label{eq:mixed-projector-explicit}
\end{equation}
Since
\begin{equation}
    n_u=-\frac{M}{2},
    \qquad
    n_\phi=-\frac{u}{2}\partial_\phi M,
\end{equation}
its two independent components are
\begin{equation}
    \Pi_u{}^a
    =
    \left(
        \frac{M}{2},1,0
    \right),
    \qquad
    \Pi_\phi{}^a
    =
    \left(
        \frac{u}{2}\partial_\phi M,0,1
    \right),
    \label{eq:projector-components-explicit}
\end{equation}
where the entries refer to the ordering \((r,u,\phi)\). A convenient
dual projector is
\begin{equation}
    \Pi_a{}^u=(0,1,0),
    \qquad
    \Pi_a{}^\phi=(0,0,1),
    \qquad
    \Pi_a{}^i\Pi_j{}^a=\delta_j{}^i.
\end{equation}
The projected derivative directions are therefore
\begin{equation}
    E_u
    \equiv
    \Pi_u{}^a\partial_a
    =
    \partial_u+\frac{M}{2}\partial_r,
    \qquad
    E_\phi
    \equiv
    \Pi_\phi{}^a\partial_a
    =
    \partial_\phi
    +\frac{u}{2}\partial_\phi M\,\partial_r .
    \label{eq:projected-derivative-directions}
\end{equation}
Notice that these two vector fields commute,
\begin{equation}
    [E_u,E_\phi]=0,
\end{equation}
as expected for the coordinate basis induced on the hypersurfaces
\(r-uM/2=\mathrm{constant}\).

The nonvanishing bulk Christoffel symbols relevant for the projected
connection are
\begin{align}
    \Gamma^r{}_{r\phi}
    &=
    \Gamma^r{}_{\phi r}
    =
    \frac{A}{r},
    &
    \Gamma^r{}_{u\phi}
    &=
    \Gamma^r{}_{\phi u}
    =
    -\frac{1}{2}\partial_\phi M,
    \nonumber\\[4pt]
    \Gamma^r{}_{\phi\phi}
    &=
    rM-\partial_\phi A-\frac{A^2}{r},
    &
    \Gamma^u{}_{\phi\phi}
    &=
    r,
    \nonumber\\[4pt]
    \Gamma^\phi{}_{r\phi}
    &=
    \Gamma^\phi{}_{\phi r}
    =
    \frac{1}{r},
    &
    \Gamma^\phi{}_{\phi\phi}
    &=
    -\frac{A}{r}.
    \label{eq:relevant-bulk-christoffel}
\end{align}

The connection induced on the hypersurface is obtained by projecting
the bulk Levi--Civita connection. In the basis \((E_u,E_\phi)\), its
coefficients are
\begin{equation}
    \widehat{\Gamma}^{k}{}_{ij}
    =
    \Pi_a{}^k\Pi_i{}^b
    \left(
        \partial_b\Pi_j{}^a
        +
        \Gamma^a{}_{bc}\Pi_j{}^c
    \right).
    \label{eq:induced-connection-explicit}
\end{equation}
The derivatives of the radial components of the projectors contribute
only in the radial direction and are therefore annihilated by
\(\Pi_a{}^u\) and \(\Pi_a{}^\phi\). Nevertheless, the nonzero radial
component of \(\Pi_\phi{}^a\) modifies the projected Christoffel terms.
One obtains
\begin{equation}
    \widehat{\Gamma}^{u}{}_{\phi\phi}
    =
    r,
    \qquad
    \widehat{\Gamma}^{\phi}{}_{u\phi}
    =
    \widehat{\Gamma}^{\phi}{}_{\phi u}
    =
    \frac{M}{2r},
    \qquad
    \widehat{\Gamma}^{\phi}{}_{\phi\phi}
    =
    \frac{\frac{u}{2}\partial_\phi M-N}{r}.
    \label{eq:induced-connection-components}
\end{equation}
Consequently,
\begin{equation}
    \widehat{\Gamma}^{i}{}_{iu}
    =
    \frac{M}{2r},
    \qquad
    \widehat{\Gamma}^{i}{}_{i\phi}
    =
    \frac{\frac{u}{2}\partial_\phi M-N}{r}.
    \label{eq:connection-traces}
\end{equation}

The change of normal also changes the projected Weingarten tensor and
hence the finite-\(r\) stress tensor. It is therefore useful to
distinguish the quantity naturally selected by the new normal,
\begin{equation}
    \mathcal{P}
    \equiv
    \frac{M}{16\pi G_N},
    \qquad
    \mathcal{K}
    \equiv
    \frac{N}{8\pi G_N},
    \label{eq:P-Ktilde-definitions-explicit}
\end{equation}
% from the quantity constructed from the metric coefficient \(A\),
% \begin{equation}
%     \mathcal{K}
%     \equiv
%     \frac{A}{8\pi G_N}
%     =
%     \mathcal{K}
%     +
%     u\,\partial_\phi\mathcal{P}.
%     \label{eq:K-Ktilde-relation}
% \end{equation}

Using the same Weingarten-tensor convention as in the main text, the
mixed stress-tensor components take the form
\begin{align}
    T_u{}^u
    &=
    -\frac{\mathcal{P}}{r}
    +\frac{\partial_\phi\mathcal{K}}{r^2}
    -\frac{
        u\,\partial_\phi M\,\mathcal{K}
    }{2r^3},
    \label{eq:Tuu-PK}
    \\[4pt]
    T_\phi{}^u
    &=
    -\frac{\mathcal{K}}{r},
    \label{eq:Tphiu-PK}
    \\[4pt]
    T_u{}^\phi
    &=
    \frac{M\mathcal{K}}{2r^3},
    \label{eq:Tuphi-PK}
    \\[4pt]
    T_\phi{}^\phi
    &=
    0.
    \label{eq:Tphiphi-PK}
\end{align}

For a mixed-index tensor, the projected covariant divergence is
\begin{equation}
    \widehat{\mathcal{D}}_i T_j{}^i
    =
    E_i T_j{}^i
    +
    \widehat{\Gamma}^{i}{}_{ik}T_j{}^k
    -
    \widehat{\Gamma}^{k}{}_{ij}T_k{}^i .
    \label{eq:mixed-tensor-divergence}
\end{equation}

We first consider the \(j=u\) component. Using
\eqref{eq:projected-derivative-directions} and
\eqref{eq:Tuu-PK}--\eqref{eq:Tphiphi-PK}, one obtains
\begin{align}
    \widehat{\mathcal{D}}_i T_u{}^i
    &=
    E_u T_u{}^u
    +
    E_\phi T_u{}^\phi
    +
    \widehat{\Gamma}^{i}{}_{ik}T_u{}^k
    -
    \widehat{\Gamma}^{k}{}_{iu}T_k{}^i
    \nonumber\\[4pt]
    &=
    -\frac{\partial_u\mathcal{P}}{r}
    +O(r^{-2}).
    \label{eq:divergence-u-finite-r}
\end{align}
Thus the radial and connection terms again do not modify the leading
boundary equation.

For the \(j=\phi\) component,
\begin{align}
    \widehat{\mathcal{D}}_i T_\phi{}^i
    &=
    E_u T_\phi{}^u
    +
    E_\phi T_\phi{}^\phi
    +
    \widehat{\Gamma}^{i}{}_{ik}T_\phi{}^k
    -
    \widehat{\Gamma}^{k}{}_{i\phi}T_k{}^i
    \nonumber\\[4pt]
    &=
    E_u T_\phi{}^u
    +
    \widehat{\Gamma}^{i}{}_{iu}T_\phi{}^u
    -
    \widehat{\Gamma}^{\phi}{}_{u\phi}T_\phi{}^u
    -
    \widehat{\Gamma}^{u}{}_{\phi\phi}T_u{}^\phi,
    \label{eq:divergence-phi-expanded}
\end{align}
where all terms proportional to \(T_\phi{}^\phi\) vanish. The individual
contributions are
\begin{align}
    E_u T_\phi{}^u
    &=
    \left(
        \partial_u+\frac{M}{2}\partial_r
    \right)
    \left(
        -\frac{\mathcal{K}}{r}
    \right)
    \nonumber\\
    &=
    -\frac{\partial_u\mathcal{K}}{r}
    +
    \frac{M\mathcal{K}}{2r^2},
    \label{eq:Eu-Tphiu}
    \\[6pt]
    \widehat{\Gamma}^{i}{}_{iu}T_\phi{}^u
    &=
    \frac{M}{2r}
    \left(
        -\frac{\mathcal{K}}{r}
    \right)
    =
    -\frac{M\mathcal{K}}{2r^2},
    \label{eq:trace-connection-term}
    \\[6pt]
    -\widehat{\Gamma}^{\phi}{}_{u\phi}T_\phi{}^u
    &=
    -\frac{M}{2r}
    \left(
        -\frac{\mathcal{K}}{r}
    \right)
    =
    \frac{M\mathcal{K}}{2r^2},
    \label{eq:lower-index-connection-one}
    \\[6pt]
    -\widehat{\Gamma}^{u}{}_{\phi\phi}T_u{}^\phi
    &=
    -r
    \left(
        \frac{M\mathcal{K}}{2r^3}
    \right)
    =
    -\frac{M\mathcal{K}}{2r^2}.
    \label{eq:lower-index-connection-two}
\end{align}
The terms of order \(r^{-2}\) cancel, leaving
\begin{equation}
    \widehat{\mathcal{D}}_i T_\phi{}^i
    =
    -\frac{\partial_u\mathcal{K}}{r}
    +O(r^{-2}).
    \label{eq:divergence-phi-finite-r}
\end{equation}

The renormalized boundary divergence is defined by stripping off the
universal \(1/r\) falloff,
\begin{equation}
    \mathcal{D}_i T_j{}^i
    \equiv
    \lim_{r\rightarrow\infty}
    r\,\widehat{\mathcal{D}}_i T_j{}^i .
    \label{eq:renormalized-boundary-divergence}
\end{equation}
Hence, in terms of the density naturally associated with the new normal,
\begin{equation}
    \mathcal{D}_i
    \left\langle T_u{}^i\right\rangle_\lambda
    =
    -\partial_u
    \left\langle\mathcal{P}\right\rangle_\lambda,
    \qquad
    \mathcal{D}_i
    \left\langle T_\phi{}^i\right\rangle_\lambda
    =
    -\partial_u
    \left\langle\mathcal{K}\right\rangle_\lambda .
    \label{eq:carroll-divergence-new-normal}
\end{equation}

% For the undeformed Bondi solution, \(M=M(\phi)\), so that
% \(\partial_u\mathcal{P}=0\). Using
% \eqref{eq:K-Ktilde-relation},
% \begin{equation}
%     \mathcal{K}
%     =
%     \mathcal{K}
%     -
%     u\,\partial_\phi\mathcal{P},
% \end{equation}
% and therefore
% \begin{equation}
%     -\partial_u\mathcal{K}
%     =
%     -\partial_u\mathcal{K}
%     +
%     \partial_\phi\mathcal{P}.
% \end{equation}
% Thus, on the Bondi solution space, the second equation in
% \eqref{eq:carroll-divergence-new-normal} can equivalently be written as
% \begin{equation}
%     \mathcal{D}_i
%     \left\langle T_\phi{}^i\right\rangle
%     =
%     -\partial_u
%     \left\langle\mathcal{K}\right\rangle
%     +
%     \partial_\phi
%     \left\langle\mathcal{P}\right\rangle,
%     \label{eq:carroll-divergence-explicitly-derived}
% \end{equation}
which reproduces the standard flat-space Carrollian conservation
equation.

% There is, however, an important qualification for source-dependent
% one-point functions. If \(\partial_u\mathcal{P}\neq0\), then
% \begin{equation}
%     -\partial_u\mathcal{K}
%     =
%     -\partial_u\mathcal{K}
%     +
%     \partial_\phi\mathcal{P}
%     +
%     u\,\partial_u\partial_\phi\mathcal{P}.
%     \label{eq:Ktilde-source-relation}
% \end{equation}
% Consequently, the statement that the same derivation applies unchanged
% to arbitrary source-dependent one-point functions is not automatic for
% this choice of normal. Either the boundary superrotation density should
% be taken to be
% \begin{equation}
%     \mathcal{K}
%     =
%     \frac{N}{8\pi G_N},
% \end{equation}
% in which case the natural projected equation is
% \begin{equation}
%     \mathcal{D}_i T_\phi{}^i
%     =
%     -\partial_u\mathcal{K},
% \end{equation}
% or additional assumptions on the source dependence, such as
% \(\partial_u\mathcal{P}=0\), are required in order to recover
% \begin{equation}
%     -\partial_u\mathcal{K}
%     +
%     \partial_\phi\mathcal{P}
% \end{equation}
% without an additional term.

%%%%%%%%%%%%%%%%%%%%%%%%%%%%%%%%%%%%%%%%%%%%%%%%%%%%%%%%%%%%%%%%%%%%%%%%%%%%%%%%%%%%%%%%%%%
\section{Carrollian sources at \(\mathscr I^+\)}
\label{app:carroll sources}
The generating function $W[\lambda_A,\mathcal{C}_i^j]$ is a functional of boundary sources $\lambda_A$ for the scalar field, and $\mathcal{C}_i^j$ (Carrollian data) for the gravity. The Carrollian data $\mathcal{C}_i^j$ for gravity can be written in terms of the degenerate metric, projector and the Carrollian vector. We provide an expression for the variation of these sources. First, we set up the notation \cite{Hartong:2025jpp,Hartong:2026rbr}, which will be our building blocks for the variation.\\ 

 Future null infinity is the two-dimensional Carrollian manifold
\(\mathscr I^+\simeq\mathbb R_u\times S^1_\phi\). The Carrollian data is
\begin{equation}
    \bigl(q_{ij},\,n^i,\,k_i\bigr),
    \qquad
    q_{ij}n^j=0,
    \qquad
    k_in^i=-1,
    \label{eq:carroll-data}
\end{equation}
with \(q_{ij}\) the degenerate spatial metric, \(n^i\) the null generator, and \(k_i\) the Ehresmann connection. The normalization
\(k_in^i=-1\) is the standard Carrollian one, and it makes \(n^i\)
past-directed. One can introduce the future-directed generator
\begin{equation}
    \ell^i\equiv-n^i,
    \qquad
    k_i\ell^i=+1,
    \qquad
    q_{ij}\ell^j=0,
    \label{eq:future-generator}
\end{equation}
We use \((k_i,e^a_i)\) with dual \((\ell^i,e^i_a)\). \footnote{Here we use $a,b$ for tangent/frame indices and $i,j$ are curved indices. One shouldn't get confused with other use of $a,b$ indices in this article. Here these $a,b$ indices just represent the frame indices.}
\begin{equation}
    e^a_i\ell^i=0,
    \quad
    e^i_ak_i=0,
    \quad
    e^a_ie^i_b=\delta^a_b,
    \qquad
    \delta^i{}_j=\ell^ik_j+e^i_ae^a_j,
    \qquad
    q_{ij}=\delta_{ab}e^a_ie^b_j .
    \label{eq:carroll-completeness}
\end{equation}

The horizontal projector and the pseudo-inverse metric are
\begin{equation}
    \Pi_i{}^j\equiv\delta_i^j-\ell^jk_i=\delta_i^j+n^jk_i=e^a_ie^j_a,
    \qquad
    q^{ij}\equiv\delta^{ab}e^i_ae^j_b,
    \qquad
    q^{ij}k_j=0,
    \qquad
    q^{ik}q_{kj}=\Pi_j{}^i .
    \label{eq:projector-pseudoinverse}
\end{equation}
The Carrollian volume form is \(\eta=k_i\wedge e_1\).
In two dimensions we write \(e^a_i\to e_i\), \(e^i_a\to e^i\), and the
reference frame is
\begin{equation}
    k=du,
    \qquad
    e=d\phi,
    \qquad
    \ell^i\partial_i=\partial_u,
    \qquad
    n^i\partial_i=-\partial_u,
    \qquad
    e^i\partial_i=\partial_\phi,
    \qquad
    \eta=k\wedge e=du\wedge d\phi .
    \label{eq:reference-frame}
\end{equation}
 A Carroll boost is by definition a shift of the Ehresmann connection,
\begin{equation}
    \delta_\beta k_i=\beta\,e_i,
    \qquad
    \delta_\beta n^i=\delta_\beta\ell^i=0,
    \qquad
    \delta_\beta q_{ij}=0,
    \qquad
    \delta_\beta e^i=-\beta\,\ell^i,
    \qquad
    \delta_\beta q^{ij}=-\beta\bigl(\ell^ie^j+e^i\ell^j\bigr),
    \label{eq:boost-action}
\end{equation}
so that \(q_{ij}\) and \(n^i\) are boost invariant while \(k_i\), \(e^i\) and \(q^{ij}\) are not. A Carrollian Weyl rescaling acts as
\begin{equation}
    k_i\to e^{\sigma}k_i,
    \quad
    e_i\to e^{\sigma}e_i,
    \quad
    \ell^i\to e^{-\sigma}\ell^i,
    \quad
    q_{ij}\to e^{2\sigma}q_{ij},
    \quad
    \eta\to e^{2\sigma}\eta .
    \label{eq:weyl-action}
\end{equation}
Both preserve \eqref{eq:carroll-data}. Since the Carrollian measure carries
Weyl weight two and boost weight zero, \(\Delta_A=2\) is the marginal dimension, and Weyl and boost weights enter sources and operators with opposite signs. The 3d Bondi gauge also induces the same frame as shown in the previous section.

The two-form \(dk\) is the curvature of the Ehresmann connection. It measures the failure of the horizontal distribution to be integrable. It is not boost-invariant and gives rise to the Carroll boost anomaly; see \cite{Hartong:2025jpp,Hartong:2026rbr}. To reach a frame with \(dk\neq0\), and hence to see the boost anomaly, one must relax the Bondi conditions \(g_{rr}=g_{r\phi}=0\), \(g_{ur}=-1\) and admit the most general Carroll structure at \(\mathscr I^+\).

\subparagraph{The Carrollian source \(\mathcal C_i{}^j\):-}
Because \(q_{ij}\) is degenerate, there is no inverse metric with which to raise the indices of a response conjugate to \(\delta q_{ij}\), and the pseudo-inverse \eqref{eq:projector-pseudoinverse} is boost dependent. Moreover the variations of \eqref{eq:carroll-data} are not independent:
varying the two defining conditions gives
\begin{equation}
    \delta q_{ij}\,n^j=-q_{ij}\,\delta n^j,
    \qquad
    n^i\,\delta k_i=-k_i\,\delta n^i .
    \label{eq:variation-constraints}
\end{equation}
One therefore cannot treat \(\delta q_{ij}\) and \(\delta n^i\) as
independent sources. The object that packages the constrained
variations into an unconstrained source is the mixed-index frame deformation
\begin{equation}
   \delta \mathcal C_i{}^j
    \equiv
    \delta e^A_i\,e^j_A
    =
    \delta k_i\,\ell^j+\delta e^a_i\,e^j_a
    =
    -k_i\,\delta \ell^j-e^a_i\,\delta e^j_a ,
    \label{eq:carrollian-source-definition}
\end{equation}
the degenerate analogue of \(\delta g_{ik}g^{kj}\); the last two forms agree
because \(\delta(e^A_ie^j_A)=0\). Eliminating the constrained components
using \eqref{eq:variation-constraints} and decomposing in the frame yields
the covariant expression
\begin{equation}
   \delta \mathcal C_i{}^j
    =
    k_i\,\delta n^j
    +
    \tfrac12\,q^{jk}\,\Pi_i{}^l\,\delta q_{kl}
    +
    \bigl(\Pi_i{}^k\,\delta k_k\bigr)\,\ell^j .
    \label{eq:carrollian-source-decomposition}
\end{equation}
In two dimensions, the middle term collapses to the conformal factor of
\(q_{ij}\), and \(\Pi_i{}^k\delta k_k=\beta\,e_i\) with
\(\beta=e^k\delta k_k\), so that
\begin{equation}
  \delta  \mathcal C_i{}^j
    =
    k_i\,\delta n^j
    +
    \tfrac12\bigl(q^{kl}\delta q_{kl}\bigr)\Pi_i{}^j
    +
    \beta\,e_i\ell^j .
    \label{eq:carrollian-source-2d}
\end{equation}
This is the source term used in the Ward identity in section \ref{sec:scalar-source-bms-ward}. When we contract this with the stress tensor we get total variation of the action which matches with eq (3.25) of \cite{Chandrasekaran:2021hxc}. Hence, both analyses are consistent with each other.
\subparagraph{Carrollian Weyl Ward identity:-}

Consistency of \eqref{eq:generating-functional-variation} with
\eqref{eq:weyl-action} requires the sources to carry the opposite Weyl weight
to the operators, \(\delta^{\rm W}_\sigma\lambda_A=(\Delta_A-2)\sigma
\lambda_A\). Since \(\delta \mathcal C_i{}^j=\sigma\delta_i^j\) %\vaishnavi{how do we get this relation for Cij?},\hare{do the variation, D.7} 
invariance of \(W\)
up to a local anomaly gives
\begin{equation}
    \left\langle T_i{}^i\right\rangle_{\lambda}
    +
    \sum_A(\Delta_A-2)\lambda_A\left\langle\mathcal O_A\right\rangle_{\lambda}
    =
    \mathcal A_{\rm W},
    \qquad
    \left\langle T_i{}^i\right\rangle=\mathcal S-\mathcal P .
    \label{eq:weyl-ward}
\end{equation}
The anomaly \(\mathcal A_{\rm W}\) must be a local functional of the
background Carroll data alone; a state-dependent quantity cannot be an
anomaly. Being built from Carrollian curvature
invariants, \(\mathcal A_{\rm W}\) vanishes on the frame
\eqref{eq:reference-frame}. We encourage the reader to consult the articles \cite{Hartong:2025jpp,Hartong:2026rbr} for more details on the Weyl and Carrollian boost Ward identities.
\section{CFT stress tensor}
\label{app:CFT stress tensor}
 
The induced Carroll structure on the surface \(r=R\) is
\begin{equation}
    q^{(R)}_{ij}=R^2\,q^{(0)}_{ij},
    \qquad
    n^{i}_{(R)}=n^{i}_{(0)}=-\partial_u,
    \qquad
    k^{(R)}_i=k^{(0)}_i=du ,
    \label{eq:cutoff-carroll-data}
\end{equation}
where \((q^{(0)},n_{(0)},k^{(0)})\) is the frame \eqref{eq:reference-frame}
induced at \(\mathscr I^+\). Only the degenerate metric carries radial
weight. The normal and auxiliary vectors are weighted zero under radial scaling. Consequently, the cut volume form and the Carrollian
measure scale linearly,
\begin{equation}
    \mu^{(R)}=R\,\mu^{(0)},
    \qquad
    \eta^{(R)}=R\,\eta^{(0)} .
    \label{eq:volume-form-scaling}
\end{equation}
 The source \(\mathcal C_i{}^j\) \eqref{eq:carrollian-source-definition}
carries \emph{no} radial weight at all.
\begin{equation}
    \delta e^{a}_{i}\big|_{(R)}\;e^{j}_{a}\big|_{(R)}
    =
    \bigl(R\,\delta e^{a}_{i}\big|_{(0)}\bigr)
    \bigl(R^{-1}e^{j}_{a}\big|_{(0)}\bigr)
    =
    \delta e^{a}_{i}\big|_{(0)}\;e^{j}_{a}\big|_{(0)} ,
    \label{eq:source-is-radially-inert}
\end{equation}
while the term \(\delta k_i\,\ell^j\) does not scale by
\eqref{eq:cutoff-carroll-data}. Hence
\(\mathcal C_i{}^j\big|_{(R)}=\mathcal C_i{}^j\big|_{(0)}\).
 
All of the radial dependence of the variational pairing therefore resides in the product \(\eta^{(R)}\,T_j{}^{i}\big|_{(R)}\). Since the on-shell action $\delta W= \int \eta T_j{}^i  \mathcal{C}_i^j$ is a fixed functional of the bulk solution and cannot depend on the
cutoff. Hence $\delta W$ evaluated at
\(r=R\) must be \(R\)-independent, which by
\eqref{eq:volume-form-scaling} and \eqref{eq:source-is-radially-inert} forces
\begin{equation}
    T_j{}^{i}\big|_{(R)}
    =
    \frac{1}{R}\,\mathcal T_j{}^{i}
    +
    \mathcal O\!\left(R^{-2}\right).
    \label{eq:radial-expansion-of-T}
\end{equation}
The boundary stress tensor can be given by
\begin{equation}
    \mathcal T_j{}^{i}
    \equiv
    \lim_{R\to\infty}R\;T_j{}^{i}\big|_{(R)},
    \qquad
    \delta W
    =
    \int_{\mathscr I^+}\eta^{(0)}\,
    \mathcal T_j{}^{i}\,\delta\mathcal C_i{}^j
    +\ldots
    \label{eq:boundary-stress-tensor-definition}
\end{equation}

The procedure is the exact Carrollian counterpart of the definition of the
holographic stress tensor in asymptotically anti-de Sitter spacetimes
\cite{Balasubramanian:1999re,deHaro:2000vlm}. In Fefferman-Graham gauge the
induced metric is \(h_{\mu\nu}=g_{(0)\mu\nu}/z^2\), and
\(\sqrt{-h}=z^{-d}\sqrt{-g_{(0)}}\), while the analogue of our source,
\((h^{-1}\delta h)_\mu{}^\nu\), is inert in \(z\) for exactly the reason
\eqref{eq:source-is-radially-inert}. Finiteness of the pairing then gives
\begin{equation}
    T^{\mu}{}_{\nu}\big|_{\rm BY}=z^{d}\,T^{\mu}{}_{\nu}\big|_{\rm CFT},
    \qquad
    \text{equivalently}
    \qquad
    T_{\mu\nu}\big|_{\rm BY}=z^{d-2}\,T_{\mu\nu}\big|_{\rm CFT},
    \label{eq:ads-analogue}
\end{equation}
which is the standard definition. The mixed-index stress tensor always
carries the inverse weight of the volume form, and it is the pairing rather
than either factor separately that is required to be finite.

\bibliographystyle{JHEP}
\bibliography{refs}

\end{document}